\documentclass[aps,pra,twocolumn]{revtex4-2}
\usepackage[english]{babel}
\usepackage{graphicx}
\usepackage{amssymb}
\usepackage{amsmath}
\usepackage{physics}
\usepackage{textcomp}
\usepackage{xcolor}

\newcommand{\Eq}[1]{Eq.\,(\ref{#1})}
\newcommand{\Eqs}[2]{Eqs.\,(\ref{#1}) and (\ref{#2})}

\newcommand{\Fig}[1]{Fig.\,\ref{#1}}
\newcommand{\Figs}[2]{Figs.\,\ref{#1} and \ref{#2}}
\newcommand{\Sec}[1]{Sec.\,\ref{#1}}
\newcommand{\be}{\begin{equation}}
	\newcommand{\ee}{\end{equation}}
\newcommand{\bi}{\begin{itemize}}
	\newcommand{\ei}{\end{itemize}}
\newcommand{\bea}{\begin{eqnarray}}
	\newcommand{\eea}{\end{eqnarray}}
\newcommand{\App}[1]{Appendix~\ref{#1}}
\newcommand{\Tab}[1]{Table~\ref{#1}}
\newcommand{\Onlinecite}[1]{Ref.~\cite{#1}}

\newcommand{\rp}{\ensuremath{r_{\rm p}}}
\newcommand{\fwa}{\ensuremath{w_{\theta}}}
\newcommand{\fwr}{\ensuremath{w_{r}}}
\newcommand{\epsw}{\ensuremath{\varepsilon_{\rm w}}}
\newcommand{\epsww}{\ensuremath{\epsilon_{\rm w}}}
\newcommand{\kTE}{\ensuremath{k_{\rm TE}}}
\newcommand{\kTM}{\ensuremath{k_{\rm TM}}}

\newcommand{\cH}{\mathcal{H}}
\newcommand{\cE}{\mathcal{E}}
\newcommand{\cK}{\mathcal{K}}
\newcommand{\cM}{\mathcal{M}}
\newcommand{\cN}{\mathcal{N}}

\begin{document}
\title{Optical resonances in graded index spheres: A resonant-state expansion study and analytic approximations}
\author{Z. Sztranyovszky}
\author{W. Langbein}
\author{E. A. Muljarov}
\affiliation{School of Physics and Astronomy, Cardiff University, Cardiff CF24 3AA, United Kingdom}
\date{\today}

\begin{abstract}
      Recent improvements in the resonant-state expansion (RSE), focusing on the static mode contribution,  have made it possible to treat transverse-magnetic (TM) modes of a spherically symmetric system with the same efficiency as their transverse-electric (TE) counterparts. We demonstrate here that the efficient inclusion of static modes in the RSE results in its quick convergence to the exact solution regardless of the static mode set used. We then apply the RSE to spherically symmetric systems with continuous radial variations of the permittivity. We show that in TM polarization, the spectral transition from whispering gallery to Fabry-P\'erot modes is characterized by a peak in the mode losses and an additional mode as compared to TE polarization. Both features are explained quantitatively by the Brewster angle of the surface reflection which occurs in this frequency range. Eliminating the discontinuity  at the sphere surface by using linear or quadratic profiles of the permittivity modifies this peak and increases the Fabry-P\'erot mode losses, in qualitative agreement with a reduced surface reflectivity. These profiles also provide a nearly parabolic confinement for the whispering gallery modes, for which an analytical approximation using the Morse potential is presented. Both profiles result in a reduced TE-TM splitting, which is shown to be further suppressed by choosing a profile radially extending the mode fields. Based on the concepts  of ray optics, phase analysis of the secular equation, and effective quantum-mechanical potential for a wave equation, we have further developed a number of useful approximations which shed light on the physical phenomena observed in the spectra of graded-index systems.
\end{abstract}

\maketitle

\section{Introduction}
\label{s:introduction}

Modeling inhomogeneous optical resonators is challenging as generally a simple analytic solution is not available. A special case are spherically symmetric systems, having an inhomogeneity, for example in the permittivity, only dependent on the radius. Examples can be found in core-shell systems which allow highly directional scattering~\cite{LiuACSN12}, when modeling surface contamination on a sphere due to diffusion~\cite{WyattPR62} or high pressure~\cite{ChowdhuryJOSAA91}, or when model biological cells~\cite{HuangPRE03}. Graded index profiles can be used to engineer the cancellation of electric and magnetic dipole excitation which reduces the visibility of small particles at certain wavelengths~\cite{ShalashovTAP16}. Graded index profiles can also lead to reduced splitting between transverse-electric (TE) and transverse-magnetic (TM) modes which enhances sensitivity to chiral materials.

The scattering properties of systems with graded permittivity have been studied in the literature using various approximate methods. In the multilayer approach (also referred to as stratified medium method), the graded index profile is approximated by a piecewise constant function, describing the system by homogeneous regions comprising a core covered by a sequence of shells~\cite{WaitASRB62, KaiAO94}. In the short wavelength limit, a Debye series expansion for the scattered field was used~\cite{LockJQSRT17}, and in the long wavelength limit a Born approximation~\cite{AlbiniJAP62} or a dipole limit~\cite{ShalashovTAP16} were applied to dispersive systems with complex permittivity. Furthermore, the dipole moment of dielectric spherical particles with power law radial profiles of the permittivity was calculated in the electrostatic limit \cite{DongPRB03}.
A generalized scattered field formulation developed in~\cite{WyattPR62} requires solving scalar Schr\"odinger-like equations, similar to the scalar wave equations solved in this work. To study the electromagnetic (EM) modes, first and second-order perturbation methods were developed~\cite{LaiPRA90} and applied to deformations of a homogeneous sphere~\cite{Leung_Pang_1996}. Whispering gallery (WG) modes in both TE and TM polarizations were studied in~\cite{ChowdhuryJOSAA91} for small inhomogeneous perturbations of the surface layer of a sphere. In that approach, the modes were found in the complex frequency plane based on the expansion coefficients of the generalized scattered field, and the secular equations were solved numerically using a Runge-Kutta method. The effect of a linearly changing permittivity profile was investigated in~\cite{IlchenkoJOSAA03} for high-frequency TE modes in large spheres, using Airy functions as an approximate solution to the corresponding scalar problem.  Finally, in~\cite{LaquerbeAWPL17}, a resonant mode of a sphere was treated in the electrostatic limit, for a negative and frequency dependent permittivity, described by an undamped (i.e. non-absorbing) Drude model, with radial dependencies of the permittivity and the electric field approximated by polynomials.

Here we will use the resonant-state expansion (RSE) to study the modes of graded index spherical resonators. The RSE is a rigorous theoretical method in electrodynamics for calculating the resonant states (RSs) of an arbitrary open optical system~\cite{MuljarovEPL10}. Using the RSs of a basis system, which can be chosen to be analytically solvable, such as a homogeneous dielectric sphere in vacuum, the RSE determines the RSs of the target system by diagonalizing a matrix equation containing a perturbation. This perturbation is defined as the difference between the basis and target systems and is expressed as a change of the permittivity and permeability distributions with respect to the basis system~\cite{MuljarovOL18}.

For a general perturbation, one needs to include in the RSE static modes~\cite{DoostPRA14,LobanovPRA19} alongside the RSs via a Mittag-Leffler (ML) representation of the dyadic Green's function. Note that the latter is at the heart of the RSE approach. Recently, the RSE has been reformulated~\cite{MuljarovPRA20}, in order to eliminate static modes, and the illustrations provided for perturbations of the size and refractive index of a homogeneous sphere show a significantly improved convergence compared to the original version of the RSE~\cite{LobanovPRA19}. The approach~\cite{MuljarovPRA20} has also proposed, though without providing illustrations, another quickly convergent version of the RSE, the one which keeps static modes in the basis.

In this paper, we consider both versions of the reformulated RSE, with and without static modes, demonstrating a similar efficiency for both. Using the RSE, we then investigate spherically symmetric inhomogeneous systems, with graded permittivity profiles. The RSs in such systems are still split into TE and TM polarizations, and are characterized by the azimuthal ($m$) and angular  ($l$) quantum numbers.  Importantly, while some graded profiles are approximately solvable analytically, the RSE can treat arbitrary perturbations and finds all the RSs of the system within the spectral coverage of the basis used, thus generating a full spectrum. This allows us to identify some prominent features in spectra, such as the quasi-degeneracy of modes and the Brewster angle phenomenon, and ultimately to engineer the shape of the spectrum via changing the permittivity profile. 

The paper is organized as follows. In \Sec{s:sphere} we study the TE and TM RSs of a homogeneous sphere, using a qualitative ray picture of light propagation and a more rigorous phase analysis of the secular equations describing the light eigenmodes, both approaches introducing several useful approximations. In \Sec{s:graded index} we briefly describe the RSE method and its optimizations used here for calculating the RSs of a graded index sphere. We then recap the analogy between wave optics and quantum mechanics, by introducing a radial Schr\"odinger-like wave equation containing an effective potential.  The RSs of a sphere with linear and quadratic radial permittivity profiles eliminating the discontinuity at the sphere surface are then discussed, and an approximate analytical solution using Morse's potential is presented. In \Sec{s:tetm_degeneracy} we investigate the TE-TM RS splitting and its reduction for graded index profiles. Details of calculations are provided in Appendices, including a comparison of the performance of the two optimized versions of the RSE, with and without elimination of static modes.

\section{Homogeneous sphere}
\label{s:sphere}

\begin{figure}[t]
	\includegraphics[width=1\linewidth]{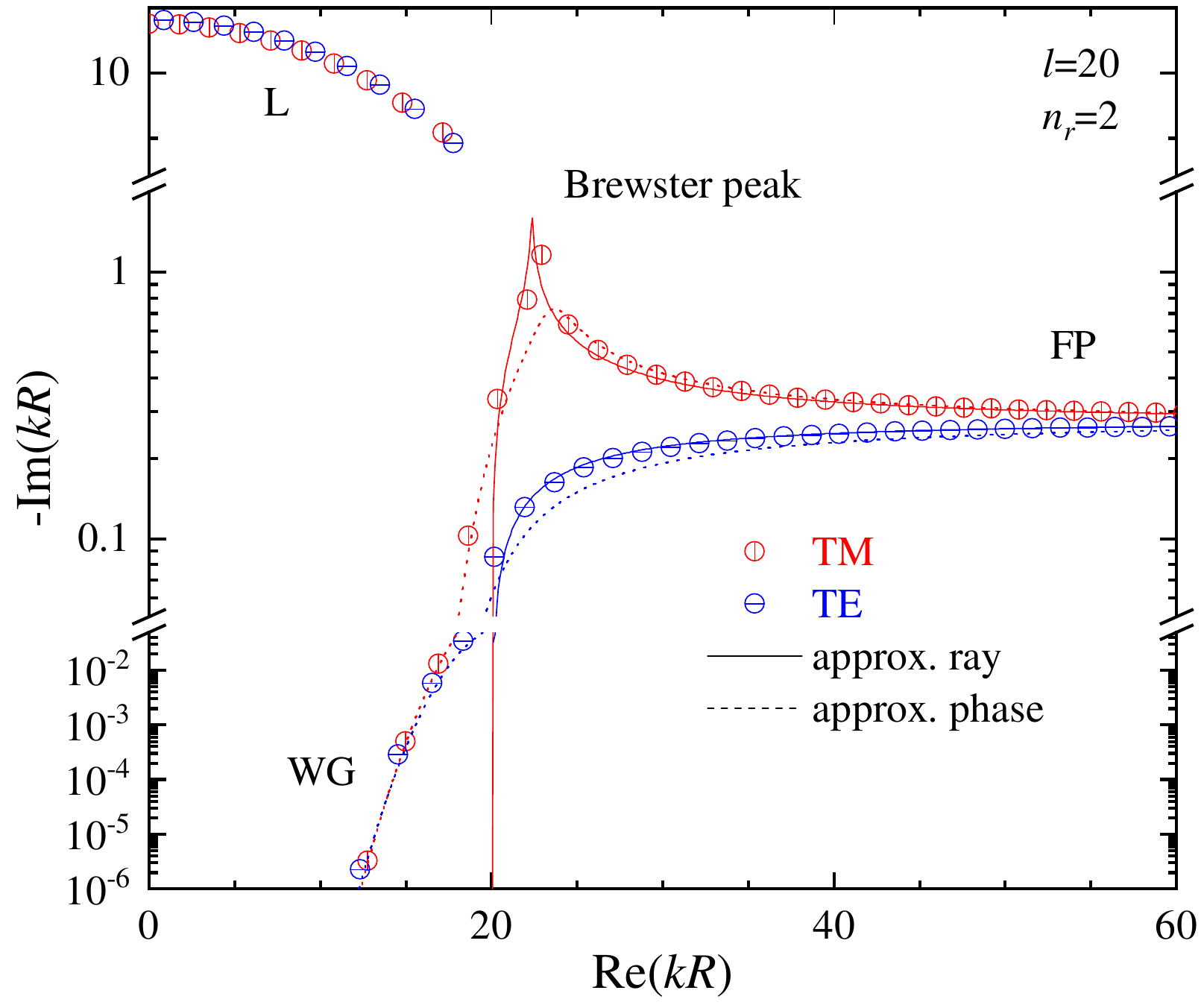}
	\caption{Wavenumbers of the TE and TM RSs of a homogeneous sphere in vacuum, with a refractive index of $n_r=2$ and an angular momentum quantum number of $l=20$. Solid and dashed lines are the approximations to the imaginary part of the wavenumbers, given by \Eqs{eq:imk}{eq:kpp}, respectively. }
\label{f:sphere}
\end{figure}

Figure ~\ref{f:sphere} shows the spectrum of the RSs of a homogeneous dielectric sphere in vacuum in the complex wavenumber plane, for a refractive index of the sphere of $n_r=2$ and an angular momentum quantum number of $l=20$. The RS wavenumbers are found by solving the secular equation, see \Eq{eq:secular_equation} in subsection ~\ref{ss:phase_analysis}.
Here, $k=\omega/c$ is the wavenumber in vacuum, $\omega$ is the light angular frequency and $c$ is the speed of light in vacuum. Only Re\,$k\ge0$ is shown, noting that RSs come in pairs with both signs of the real part of their wavenumber. The spectrum consist of TE and TM modes which appear in alternating order, with one exception related to the Brewster's angle phenomenon, as discussed below. The RSs of a sphere can be divided into three groups: leaky (L) modes, WG modes, and Fabry-P\'erot (FP) modes.

Physically, all of them are formed as a results of light quantization in the system which is provided by a constructive interference of electromagnetic (EM) waves multiply reflected from the sphere surface, but this effect is more prominent for WG and FP modes.

L modes typically have very low quality factors (Q factors) and their EM fields are located mainly outside the sphere. The number of L modes is exactly $l$ in TE and $l-1$ in TM polarization, although the Brewster mode discussed later can be regarded as a hybrid L-FP mode, so that one could say that the number of L modes is effectively the same in both polarization. L modes arrange around the origin in the complex wavenumber plane, forming a roughly semicircular arc.

WG modes are formed due to the total internal reflection and therefore have wavenumbers with $|{\rm Re}\,k|<l/R$, as discussed below. The number of WG modes is increasing with $n_r$ and $l$. The Q factor of the fundamental WG mode is increasing exponentially with $l$, and values of up to $10^{10}$, only limited by material properties, have been demonstrated experimentally \cite{VernooyOL98}. The EM field of the WG modes is concentrated inside the sphere close to the surface.

FP modes of a sphere have moderate Q factors and are named for their similarity to the original FP modes \cite{PerotAJ1899} of a double-mirror planar resonator. In fact, at large frequency, the FP modes of a sphere approach the limit of an equidistant spectrum of a dielectric slab, with all the eigenfrequencies having the same imaginary part~\cite{MuljarovEPL10}. The number of FP modes is countable infinite. Their EM fields are distributed within the sphere, avoiding the centre due to the non-zero angular momentum ($l>0$). The FP modes are spectrally separated from the WG modes by the critical angle of the total internal reflection, as discussed in more depth below.

The arrangement of the RSs in \Fig{f:sphere} is overall similar in the TE and TM polarizations. The imaginary part of their wavenumbers approaches the same high frequency asymptote, albeit from opposite sides. Additionally, there is a peak in the imaginary part of the TM RS wavenumbers near the transition region from WG to FP modes, which occurs around the Brewster angle in the ray picture of light propagation, and we therefore refer to it as a Brewster peak. At this peak, an additional TM mode is formed, breaking the otherwise alternating order of TE and TM RSs.

Below we discuss and analyze the spectrum of the RSs of a sphere in more detail, using two different approaches: the ray picture and a phase analysis. Both approaches provide some useful approximations for the mode positions and linewidths and offer an intuitive understanding of the origin and properties of the RSs of a sphere.

\subsection{Ray picture: Brewster's phenomenon  and total internal reflection}
\label{ss:brewster}

\begin{figure}[t]
	\centering
	\includegraphics[width=1\linewidth]{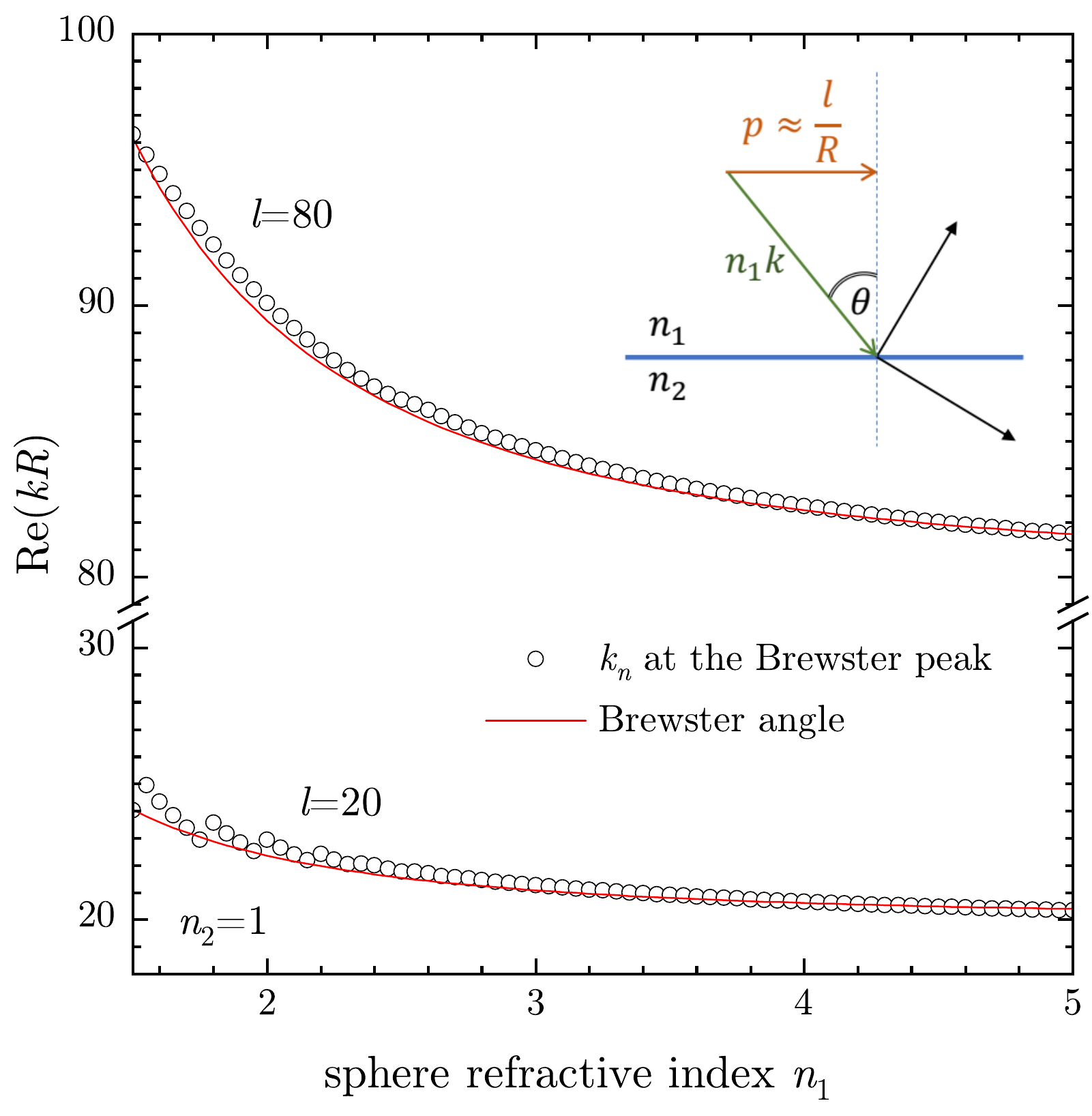}
	\caption{Real part of the wavenumber of the TM RS at the Brewster peak (circles) and function of the sphere refractive index $n_1=n_r$, for $l=20$ and $l=80$, compared with the ray optics approximation \Eq{eq:brewster} (lines). Inset: sketch of the  ray geometry at the boundary.}
	\label{f:brewster}
\end{figure}

To understand the observation of the Brewster peak in the spectrum of the RSs, we recall that increasing the angle of light incidence $\theta$ at a planar interface between two media, the Fresnel reflection coefficient for TM (aka p) polarized light passes through zero, changing its sign at the Brewster angle~\cite{GriffithsBook17}. The same occurs at the surface of a sphere in the ray picture, which is valid in the limit of wavelengths much smaller than the surface curvature. This local geometry is illustrated in the inset of \Fig{f:brewster}. The magnitude of the incident wave vector is $n_1 k$, where $n_1$ is the refractive index of the corresponding medium, i.e. that the sphere, $n_1=n_r$. Since the angular momentum $l$ gives the number of wave periods along one circumference $2\pi R$, the wave vector component $p$ parallel to the surface is determined by $2\pi l=2\pi R p$, so that $p = l/R$. With simple trigonometry we can see that $\sin \theta = p / (n_1 k)$. The Brewster angle $\theta_b$ is determined by $\tan \theta_b = n_2/n_1 $, so that for a sphere in vacuum ($n_2 =1$) the wavenumber corresponding to the Brewster angle is given by
\begin{equation} \label{eq:brewster}
    k_b = \frac{l}{R} \sqrt{\frac{1}{n_1^2} + 1}\,.
\end{equation}
At this angle, the reflectivity vanishes. This would correspond to a divergence of the imaginary part of the RS wavenumber for an ideal planar geometry. Here instead  it is kept finite due to the finite curvature of the surface and the RS discretization, resulting in the Brewster peak.

In \Fig{f:brewster} we compare \Eq{eq:brewster} with the real part of the Brewster mode (the TM mode at the Brewster peak in the spectrum), for $l=20$ and $l=80$, both showing good agreement. With increasing $n_1$ the RSs are packed more densely in the complex $k$ plane, so that the discretization does not result in significant deviations. At the same time, the light wavelength within the sphere $2\pi/(n_1k)$ decreases with $n_1$, thus improving the validity of the ray picture.

The Brewster mode can also be associated with the leaky branch. In fact, as $n_1$ increases, the Brewster peak in the spectrum is getting sharper, so that the Brewster mode is taking a significantly larger imaginary part of the wavenumber compared to the neighboring FP modes and is thus getting more isolated from them, at the same time approaching the edge of the leaky branch. Indications of this can be seen in \Figs{f:sphere_app}{f:strength_perturbation} in the Appendix. We also note that for high $l$, the Brewster peak can be shifted further into the FP spectral region. This happens because the Brewster angle $\theta_b$ is always smaller than the critical angle $\theta_c$ of the total internal reflection. The latter determines the point in the spectrum separating WG from FP modes and can be evaluated in a similar way, leading to $k_c = {l}/{R}$. Comparing it with \Eq{eq:brewster}, we see that as $l$ increases or $n_1$ decreases, the difference $k_b-k_c$ is getting larger, so that the corresponding region in the spectrum, between the critical and the Brewster angles, can accommodate more RSs.

The ray picture is also useful for understanding the imaginary part of the FP mode wavenumbers. Assuming the reflectivity amplitude $r_P$ at the sphere surface in polarization $P$ is given by the corresponding Fresnel coefficient, we equate it to the ratio of the field amplitude before and after each reflection. This ratio is in turn given by the temporal decay of the field, $|r_P|=\exp(-t/\tau)$, where $t$ is the time between consecutive reflections and $\tau$ is the mode decay time which is given by the imaginary part of its eigenfrequency, $1/\tau=-\Im(kc)$. At the same time, the optical path length across the sphere between two reflections is given by $L = 2Rn_r\cos\theta$. Finally, using the fact that    $t=L/c$ and taking the logarithm of the reflectivity results in

\begin{equation}
\label{eq:imk}
\Im k=\frac{\ln|r_P|}{2Rn_r\cos\theta},
\end{equation}
where the Fresnel coefficient $r_P$ depends on the angle of incidence $\theta$ and the refractive index of the sphere $n_r$. The expression is valid up to the critical angle $\theta_c$ of total internal reflection, at which $\ln|r_P|=0$. The values obtained according to \Eq{eq:imk} are shown in \Fig{f:sphere} as solid lines. We can see a good agreement for both polarizations, including the Brewster peak and the asymptotic value for FP modes, evaluated to $-0.27465/R$ for $n=2$ and $\theta=0$, which again validates the ray optics interpretation of the RS properties. The WG modes are located in the total internal reflection region of the spectrum where \Eq{eq:imk} is not applicable -- their non-vanishing imaginary parts are the result of the finite curvature of the sphere making the reflection imperfect. We therefore consider in the following subsection a refined approximation (shown in \Fig{f:sphere} by dashed lines) which is based on the phase analysis of the secular equation determining the RSs.

\subsection{Phase analysis: Mode positions and linewidths}
\label{ss:phase_analysis}

The secular equation determining the RS eigen wavenumber $k_n$ of a non-magnetic homogeneous sphere of radius $R$ with vacuum outside is given by~\cite{MuljarovPRA20}
\begin{equation}
\label{eq:secular_equation}
\frac{J'(n_r k_n R)}{J(n_r k_n R)} = \frac{1}{\beta} \frac{H'(k_n R)}{H(k_n R)},
\end{equation}
where $\beta = n_r$ ($\beta=n_r^{-1}$) for TE (TM) polarization. Here $J(x) = xj_l(x)$ and $H(x)=xh^{(1)}_l(x)$, with $j_l$ and $h^{(1)}_l$ being, respectively, the spherical Bessel function and Hankel function of first kind, and primes mean the first derivatives of functions with respect to their arguments. For $|z| \gg l$, we can approximate the left hand side of \Eq{eq:secular_equation} as~\cite{SehmiPRB20}
\begin{equation}
\label{eq:tan_approximation}
\frac{J'(z)}{J(z)} \approx -\tan(z-\frac{l+1}{2} \pi)\,.
\end{equation}
It is therefore useful to introduce the following two phase functions:
\begin{equation}
\label{eq:psi}
\Psi(k) = \atan(-\frac{J'(n_r kR)}{J(n_r kR)})
\end{equation}
and
\begin{equation}
\label{eq:phi}
\Phi(k) =\atan\left(-\frac{1}{\beta} \frac{H'(kR)}{H(kR)}\right)\,.
\end{equation}
Substituting them into \Eq{eq:secular_equation} yields
\be
\label{eq:sec_equ}
\Psi(k_n)=\Phi(k_n)+n\pi\,,
\ee
where $n$ is an arbitrary integer. For real $k$, it can be seen that $\Psi(k)$ is a real monotonous function (on a selected Riemann sheet), and according to \Eq{eq:tan_approximation} becomes linear  at large $k$. At the same time, $\Phi(k)$ is complex even for real $k$, and its real part varies between $\pi/2$ and 0 monotonously (non-monotonously) with $k$  for TE (TM) polarization. All three functions, $\Psi(k)-n\pi$, and Re\,$\Phi(k)$ for TE and TM polarizations, are plotted in \Fig{f:phase} in \App{a:phase_analysis}, which allows a graphical solution of \Eq{eq:sec_equ}. Namely, the intersections of the curves for $\Psi(k)-n\pi$ and Re\,$\Phi(k)$ determine the approximate positions of the modes in spectra. More rigorously, separating the real and the imaginary parts of the wavenumber, $k_n=k'_n+ik_n''$, the mode positions in spectra, $k_n'$, are given by
\begin{equation}
\label{eq:phipsi}
\Psi(k_n') - n\pi \approx {\rm Re}\,\Phi(k_n')\,,
\end{equation}
whereas $k_n''$, determining the mode linewidths, by
\begin{equation}
\label{eq:kpp}
k_n''\approx \frac{1}{n_rR}\, {\rm Im}\,\Phi(k_n')\,,
\end{equation}
in accordance with the asymptotic behaviour \Eq{eq:tan_approximation}.

The approximation \Eq{eq:kpp} for the mode linewidth is illustrated in \Fig{f:sphere} by dashed lines, demonstrating a good agreement for WG and FP modes. While it is less accurate than \Eq{eq:imk} for most FP modes, it provides a suited approximation for the WG modes, where the latter fails. The accuracy provided by this approximation improves as the refractive index $n_r$ of the sphere increases, as seen in \Fig{f:sphere_app} in \App{a:phase_analysis}. Compared to Eq. (1.1) of \cite{LamJOSAB92}, here \Eq{eq:phipsi} is not an explicit expression for mode position, and the approximation \Eq{eq:kpp} is less accurate than Eq. (1.3) of \cite{LamJOSAB92}, but the graphical solution (\Fig{f:phase}) provides intuition into the emergence of the modes and the difference between the TE and TM polarizations.

Using the above phase analysis, one can also obtain an analytic approximation for the RSs wavenumbers in the large frequency limit, $n_rkR \gg l$. Using the fact that $\tan \Phi(k) \to-i/\beta$ at $k\to\infty $ and the asymptotic behaviour of  $\Psi(k)$ given by \Eq{eq:tan_approximation}, one can evaluate
\begin{align}
\label{eq:mode_approximate}
\begin{split}
k^{\rm TE}_n &\approx \frac{1}{2n_r R}\left[(2n+l+1)\pi - i\ln{\frac{n_r + 1}{n_r -1}}\right]\,,\\
k^{\rm TM}_n &\approx \frac{1}{2n_r R}\left[(2n+l+2)\pi - i\ln{\frac{n_r + 1}{n_r -1}}\right]\,,
\end{split}
\end{align}
where the integer $n$ can be used to number the RSs. For a full derivation of \Eq{eq:mode_approximate}, see \App{a:phase_analysis}.

The RS wavenumbers given by the approximation \Eq{eq:mode_approximate} are identical to those of a homogeneous slab at normal incidence~\cite{MuljarovEPL10}. The latter are in turn consistent with \Eq{eq:imk} used for the normal incidence reflection, which gives $\Im(kR)=\ln[(n_r-1)/(n_r+1)]/(2n_r)$, as in \Eq{eq:mode_approximate}.
At non-normal incidence, the TE and TM FP modes of a slab asymptotically converge to each other in pairs, as shown in \Fig{f:homogeneous_slab} in \App{a:homogeneous_slab}. The planar system gives rise to both even and odd modes (using the parity of the electric or magnetic field), with odd TE modes converging to even TM modes at large frequencies, and vise versa. 
In the sphere, however, there are no even modes, as required by the finiteness of the EM field at the origin (as in any other point in space). Then, by removing the even modes from the slab spectra we obtain the alternating nature of the FP modes,  which is exactly what we see in the analytic approximation \Eq{eq:mode_approximate} and in the spectrum of the sphere presented in \Fig{f:sphere}.

\section{Graded index spheres}
\label{s:graded index}

In this section we study, using the RSE, the RSs in spherically symmetric non-magnetic systems with graded permittivity profiles. A particularly interesting situation is reached by removing discontinuities of the permittivity. Here we study cases where the discontinuity is removed either only in the permittivity (linear case) or both in the permittivity and its derivative (quadratic case), and compare both cases with each other and with the constant permittivity profile studied in \Sec{s:sphere}. We note that removing discontinuities of the refractive index yields broadband anti-reflecting coatings in planar dielectric layers~\cite{HedayatiM16}. For the WG modes, we introduce  a radial Schr\"odinger-like wave equation containing an effective potential, compare potentials and mode properties in all three cases, and provide an analytical approximation based on the Morse potential.

\subsection{Calculating the RSs via the RSE}

It is straightforward to use the RSE for calculating the RSs of a graded index sphere. The difference in the permittivity between the target system (a graded index sphere) and the basis system (a constant index sphere) is treated as a perturbation, and the RSs of the constant index sphere serve as a basis for the RSE. The EM fields of the RSs of the target system are expanded into the basis RSs, and the expansion coefficients and the RS wavenumbers of the target system are found by solving a linear eigenvalue problem, see \Eq{RSE-gen} in \App{a:RSE}. This eigenvalue problem of the RSE contains as input the RS wavenumbers of the basis system and the matrix elements of the perturbation. For spherically symmetric systems, TE and TM polarizations do not mix and can be treated separately in RSE as well as the RSs with different $l$ and magnetic quantum number $m$. However, the matrix elements used in the RSE for the TE and TM RSs are different, see~\cite{MuljarovPRA20} and \App{a:RSE} for details. In particular, for TM polarization, one needs to include in the basis additional functions which are required for completeness and physically describe the part of the EM field in a graded index sphere which is not divergence free. More rigorously, these functions are required to properly describe a longitudinal part of the dyadic GF related to its static pole in the ML explansion.

Previously, this problem has been treated within the RSE by introducing a complete set of static modes~\cite{LobanovPRA19}. However, even though the treatment of static modes is numerically less complex, a slow convergence versus the basis size observed in~\cite{LobanovPRA19} remained an issue. To develop quickly converging versions of the RSE, the full ML representation of the dyadic GF of a spherically symmetric system has been studied in~\cite{MuljarovPRA20}, focusing in particular on the static pole of the GF containing a $\delta$-like singularity. A quick convergence of the RSE has been achieved and demonstrated in~\cite{MuljarovPRA20} by an explicit isolation of the singularity that has allowed to avoid its direct expansion into static modes. Two ML forms of the GF have been introduced in~\cite{MuljarovPRA20}, called there ML3 and ML4, which led to slightly different versions of the RSE, both quickly convergent to the exact solution.

The quick convergence of the RSE based on ML4, with static mode elimination and suited only for a basis system in a form of a homogeneous sphere, was demonstrated in~\cite{MuljarovPRA20} on examples of both size and material (strength) perturbations of a sphere. However, the version of the RSE based on ML3, which is using explicitly a static mode set and an arbitrary spherically symmetric basis system, has not been studied so far numerically. Such a study is given in \App{a:RSE}, including a comparison with ML4, demonstrating a similar level of convergence. We show there in particular that the RSE based on ML3 and ML4  have both a quick $1/N^3$ convergence to the exact solution, where $N$ is the basis size of the RSE. Furthermore, taking three different static mode sets introduced earlier in~\cite{LobanovPRA19,MuljarovPRA20}, we show in \App{a:RSE} that the results of the RSE based on ML3 are similar for the different static mode sets previously suggested.

Let us finally note that for perturbations without discontinuities, the above mentioned optimization of the RSE might be not needed, as demonstrated in a similar approach based on eigen-permittivity modes~\cite{ChenJCP20}. However, as we are going to consider a transformation of an optical system from a homogeneous sphere, having a discontinuity, to a sphere with a continuous permittivity profile, the perturbation describing this transformation and used in RSE contains a discontinuity, both in linear and quadratic cases, and therefore the above optimization is in fact needed.

In all calculations of the RSs of the graded index spheres done in this paper, we use the RSE based on ML4, as it has a fixed number of additional basis functions in TM polarization, which is three times the number of the TM RSs included in the basis. We use the basis size (i.e. the total number of modes in the basis) of $N=800$ in both cases of linear and quadratic profiles.

\subsection{Effective potential}
\label{s:effective_potential}

To intuitively understand the properties of the RSs in graded-index optical systems, it is useful to consider the analogy between Maxwell's and Schr\"odinger's wave equations and to introduce an effective optical potential~\cite{JohnsonJOSAA93}. In spherically symmetric systems, all the components of the electric and magnetic fields can be expressed in terms of a radially dependent scalar field~\cite{MuljarovPRA20}. For TE (TM) polarization, this is the magnitude of  the electric (magnetic) field, which has only a tangential component $E(r)=\mathcal{E}(r)/r$ ($H(r)=-i\mathcal{H}(r)/r$). For non-magnetic systems, with the radial permittivity profile $ \varepsilon(r)$ and permeability $\mu(r) = 1$, the scalar field $\mathcal{E}(r)$ satisfies the following Schr\"odinger-like equation~\cite{MuljarovPRA20}
\begin{equation}
    \label{eq:radial_equation_TE}
            \left(  \dv[2]{}{r} - \frac{\alpha^2}{r^2} + k^2 \varepsilon(r)  \right) \mathcal{E}(r) = 0\,,
\end{equation}
where $\alpha=\sqrt{l(l+1)}$. In fact, assuming the particle mass $M=\hbar^2/2$, \Eq{eq:radial_equation_TE} can be interpreted as a quantum-mechanical analogue (QMA). An obvious limitation of this QMA is that $k^2$, playing the role of the complex eigenvalue for the RSs, contributes to \Eq{eq:radial_equation_TE} not the same way as the energy in Schr\"odinger's equation.
Associating  $k^2$ with the particle energy, and using the fact that $\varepsilon(r)=1$ (or a constant) outside the system, Johnson~\cite{JohnsonJOSAA93} introduced an energy-dependent effective potential, which makes the analogy with quantum mechanics no so straightforward. Here instead, we interpret \Eq{eq:radial_equation_TE} as
an equation for the {\it zero-energy} state of a particle in a  one-dimensional potential
 \begin{equation}
     \label{eq:TE_potential}
     V^{\rm TE}(r) =  - k^2 \varepsilon(r) + \frac{\alpha^2}{r^2}\,,
 \end{equation}
in which $k$ plays the role of a complex parameter of the potential. 
In this QMA, every RS of the optical system, described by the wave function $\mathcal{E}(r)$, has zero quantum-mechanical energy and potential \Eq{eq:TE_potential} used for this single state only, characterized by an individual value of $k$.

Likewise, for TM polarization, the scalar field $\mathcal{H}(r)$ satisfies an equation~\cite{MuljarovPRA20}
\begin{equation}
    \label{eq:radial_equation_TM}
            \left( - \frac{1}{\varepsilon(r)} \dv[]{\varepsilon}{r} \dv[]{}{r}  +  \dv[2]{}{r} - \frac{\alpha^2}{r^2} + k^2 \varepsilon(r) \right) \mathcal{H}(r) = 0\,,
\end{equation}
again, valid for a non-magnetic system described by the permittivity $ \varepsilon(r)$.
Compared to \Eq{eq:radial_equation_TE}, there is an additional term proportional to the logarithmic derivative of the permittivity, which can be included in the potential, yielding
 \begin{equation}
     \label{eq:TM_potential}
     V^{\rm TM}(r) = V^{\rm TE}(r)+\frac{\varepsilon'(r)}{\varepsilon(r)} \frac{\mathcal H'(r)}{\mathcal H(r)}\,,
 \end{equation}
where the prime indicates the spatial derivative. The second term in \Eq{eq:TM_potential} is analyzed and discussed in more depth in \Sec{ss:perturbation_method}, that in particular helps understanding of the TE-TM mode splitting. Here, we only note that this term, in its present form depending on the wave function, is inconsistent with the standard definition of the potential. However, introducing a re-scaled wave function $\mathcal{\widetilde H}(r) = \sqrt{\varepsilon(r)} \mathcal{H}(r)$ brings the effective potential to the form
\begin{equation}
	\label{eq:TM_potential_alt}
	\widetilde{V}^{\rm TM}(r) = V^{\rm TE}(r)+ \frac{3}{4} \left[\frac{\varepsilon'(r)}{\varepsilon(r)}\right]^2 - \frac{1}{2} \frac{\varepsilon''(r)}{\varepsilon(r)}\,,
\end{equation}
which is now independent of the wave function, thus providing a valid QMA also for TM polarization, as detailed in \App{a:new_potential}.

Note that the radial equations (\ref{eq:radial_equation_TE}) and (\ref{eq:radial_equation_TM}) are aligned with the standard Maxwell boundary conditions requiring that $\mathcal{E}$ and $\mathcal{E}'$ are continuous in TE polarization, and $\mathcal{H}$ and $\mathcal{H}'/\varepsilon$ are continuous in TM polarization.
Clearly, any discontinuity of $\varepsilon$ results in $\mathcal{H}'$ being also discontinuous in TM polarization, which is in particular the case of a homogeneous dielectric sphere in vacuum.

\subsection{Constant permittivity}
\label{s:homogeneous_potential}

 \begin{figure*}[!tp]
	\centering
	\includegraphics[width=1\textwidth, height=0.935\textheight]{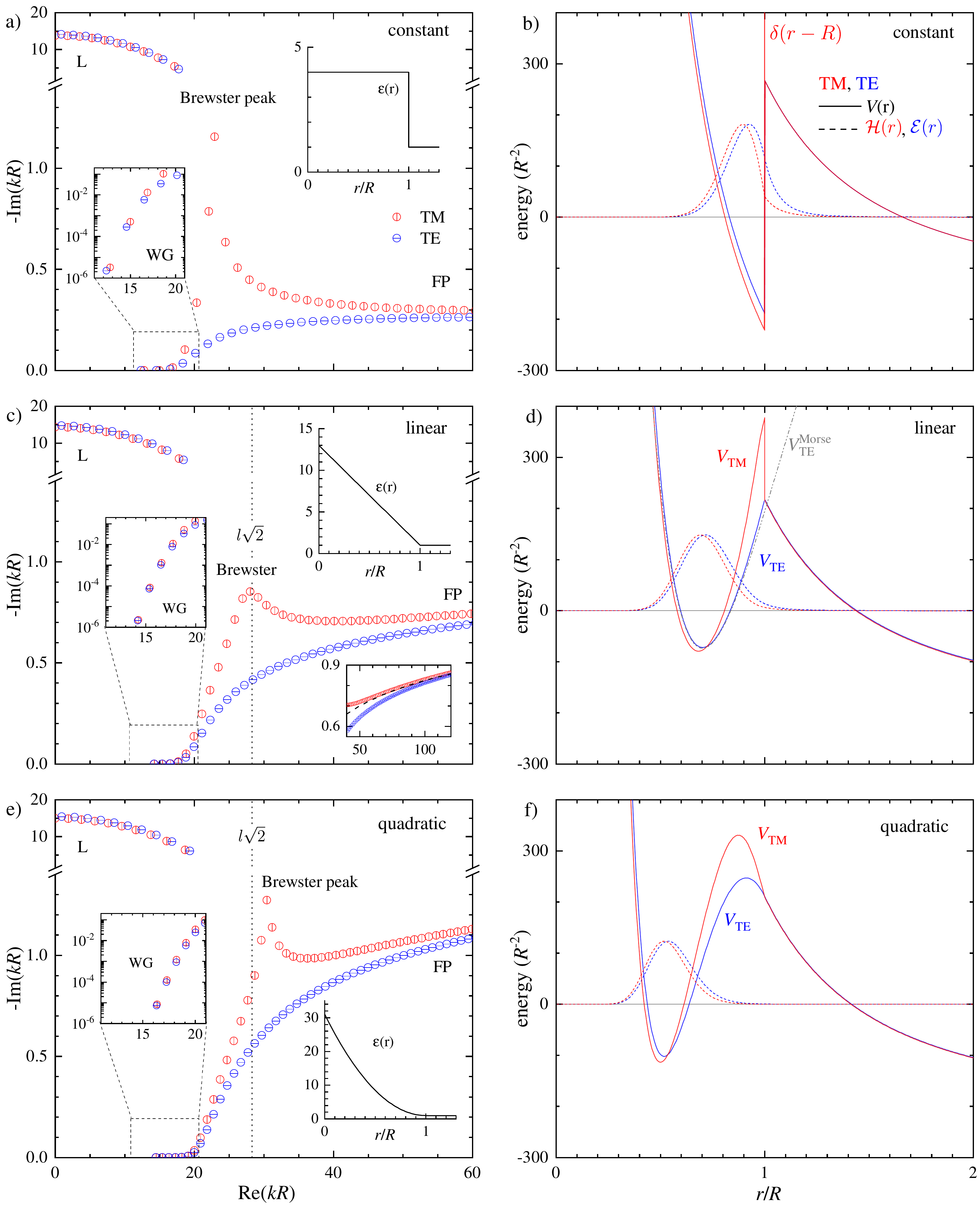}
	\caption{RSs for $l=20$, and constant (a,b), linear (c,d), and quadratic (e,f) permittivity profiles as shown in the insets. Left: RSs in the complex $k$ plane.  Right: Real part of the potential and the field of the first WG mode. The TE and TM fields are normalized to the same maximum value. $V^{TE}$ and $V^{TM}$ are given, respectively, by \Eqs{eq:TE_potential}{eq:TM_potential}.}
\label{f:modes_and_potential}
\end{figure*}

The TE and TM modes of a homogeneous sphere in vacuum, used as basis system in the RSE and described by a constant permittivity
\begin{equation}
\varepsilon(r) =1+A\theta(R-r)\,,
\end{equation}
where $\theta(x)$ is the Heaviside function and $A=n_r^2-1$,  are shown in \Fig{f:modes_and_potential}a for $n_r=2$ (note they are exactly the same as in \Fig{f:sphere}).
The fields, $\mathcal{E}(r)$ and $\mathcal{H}(r)$, and the corresponding effective potentials, given by \Eq{eq:TE_potential} and  \Eq{eq:TM_potential}, are illustrated in \Fig{f:modes_and_potential}b for the fundamental WG mode in, respectively, TE and TM polarizations. Both potentials decrease with radius due to the centrifugal term $\alpha^2/r^2$ and have similar step-like barriers at the sphere surface ($r=R$) due to the step in the permittivity. In the TM potential, there is additionally a $\delta$ function at the sphere surface due to the derivative of the permittivity, see \Eq{eq:TM_potential}. The fields are effectively confined near the sphere surface, on one side by the centrifugal term increasing towards the center of the sphere and on the other side by the refractive index step at the sphere surface. The fields have evanescent tails extending outside of the sphere, which convert at larger distances into propagating waves once the potentials become negative, and then grow exponentially due to the imaginary part of the potentials created by the complex $k$.

The optical transmission through the barrier determines the losses of the WG modes and hence the imaginary part of their wavenumbers. The height of the barrier depends on the size of the permittivity step and the angular quantum number $l$, and the transmission reduces about exponentially with $l$, thus allowing for very low mode losses~\cite{VollmerBook20}. Note that in a purely quantum-mechanical problem, having a real potential, the eigenenergy of such a state would necessarily have a finite imaginary part~\cite{Baz69} -- our potentials are however complex due to the finite imaginary part of the RS wavenumbers, though the latter is small for WG modes. Interestingly, it is the complex potential which allows the state energy in the QMA to have zero imaginary part, even though there is a finite probability for the particle to tunnel through the barrier and to escape from the system.

\subsection{Linear permittivity}
\label{ss:linear_permittivity}

We choose here a linear profile in the form
\begin{equation}
\label{eq:eps_lin}
\varepsilon(r) =1+B\theta(R-r)(1-r/R)\,,
\end{equation}
so that $\varepsilon(r)$ is a continuous function. The parameter $B$ is chosen such that the volume integral of the permittivity $\int \varepsilon(r) \dd V$ within the sphere of radius $R$ is equal to that of the homogeneous sphere with refractive index $n_r$, yielding $B=4(n_r^2-1)$. Since the basis system used in the RSE has $n_r=2$, we take here $B=12$.

The resulting RS wavenumbers calculated via the RSE are shown in \Fig{f:modes_and_potential}c. Their distribution in the complex $k$-plane is qualitatively similar to that of the homogeneous sphere. The L RSs are nearly unaffected. The WG RSs have a smaller TE-TM splitting and a quicker growth of the imaginary part of $k$ with the real part. The Brewster peak is less pronounced, broader, and is shifted towards larger values of the real part of $k$. At the sphere boundary the refractive index is approaching 1, so that using \Eq{eq:brewster} one would expect the Brewster peak to appear at around $k_b R\approx l\sqrt{2}$, which is indeed observed in the spectrum, see a dotted line in \Fig{f:modes_and_potential}c.
Note, however, that \Eq{eq:brewster} of ray optics fails in this case, as the refractive index is the same on both sides of the boundary.

The FP RS wavenumbers show a significantly larger imaginary part compared the homogeneous case. Also, it is increasing with the real part, which is qualitatively different from the homogeneous sphere, where the imaginary part of $k$ for the FP RSs is converging to a finite value with increasing the real part of $k$.  This can be understood again considering the reflection at the sphere surface.
For graded index boundaries, the reflectivity is wavelength dependent. It is proportional to the index change over one wavelength, thus proportional to $1/\Re k$ for short wavelength. An example of this can be found in \cite{YehBook05} for a segment with exponential permittivity profile. Using \Eq{eq:imk} we therefore expect $\Im k \propto \ln(\Re k)$, which is shown as a dashed line in the lower inset of \Fig{f:modes_and_potential}c, in good agreement with the high frequency asymptote of TE and TM wavenumbers.

To understand the behavior of the WG RSs, we consider the QMA, with potentials shown in \Fig{f:modes_and_potential}d. The shape of the potentials suggests that they can be approximated with the anharmonic Morse potential \cite{MorsePR29}, for which analytical solutions are known. This is explored in \App{a:morse}. A fit of the Morse potential, matching the 0th to 3rd derivative of the potential at its minimum, is shown in \Fig{f:modes_and_potential}d for the first WG mode in TE polarization. Using the analytical solutions, we find for the linear permittivity \Eq{eq:eps_lin} the following compact expression for the TE WG modes
\begin{equation}
\label{eq:knTE_approximate}
k_n^{\rm TE} \approx \frac{\alpha B}{2R(1+B)^{3/2}}\left(3+\sqrt{3}\frac{2n+1}{\alpha}-4\left(\frac{2n+1}{3\alpha}\right)^2\right)^{3/2}
\end{equation}
with the level number $n=0,1,..$. In this expression $n$ has the physical meaning of number of nodes in the field inside the resonator. The accuracy of this expression relies on a high potential barrier, providing a small tunneling (and thus small imaginary part of $k$) which is typical for WG modes. Therefore the approximation \Eq{eq:knTE_approximate} has a higher accuracy for higher $l$ and lower $n$. For $l=80$, the approximation \Eq{eq:knTE_approximate} gives $k_n$ values with the a relative error to the RSE values increasing from $10^{-5}$ for the first WG mode ($n=0$) to $10^{-2}$ for the 12th WG modes ($n=11$), as illustrated by Table~\ref{t:k_morse} in \App{a:morse}. Furthermore, \Eq{eq:knTE_approximate} creates, for $n\ll\alpha$, equidistant levels of spacing $9B/(2R\sqrt{(1+B)^3})$, resembling a harmonic oscillator.

The Morse approximation of the TM potential \Eq{eq:TM_potential_alt} for linear permittivity, and both TE and TM potentials for other spatial dependencies of the permittivity, result in non-linear simultaneous equations for $k_n^2$ as detailed in \App{a:morse}. Solving these numerically is still a lower cost compared to using the RSE or solving the radial equations (\ref{eq:radial_equation_TE}) and (\ref{eq:radial_equation_TM}) directly. The Morse approximation also provides analytical wave functions, which can be used for applying perturbation approaches like the one presented in \Sec{ss:perturbation_method} below.

\subsection{Quadratic permittivity}
\label{ss:quadratic_permittivity}

In addition to the continuity of the permittivity we can require also that its first derivative is continuous, which can be achieved by using a quadratic profile
\begin{equation}
	\varepsilon(r) =1+C\theta(R-r)(1-r/R)^2\,,
\end{equation}
where we again choose to conserve $\int \varepsilon(r) \dd V$ relative to the basis system, yielding $C=10(n_r^2-1)$, so that $C=30$ for $n_r=2$. The resulting RS wavenumbers are shown in \Fig{f:modes_and_potential}e.
The RSs change further along the same trends as seen when going from constant to linear profile. Notably, the Brewster peak is shifted further to higher wavenumbers compared to constant and linear case. 
The contrast with the surrounding is lower compared to the linear profile, creating an increased uncertainty in the position of the plane of reflection. It can also be seen from the permittivity profile that the effective radius of the sphere is reduced compared to the constant and linear cases, which results in a larger $k_b$, in accordance with \Eq{eq:brewster}.

The imaginary part of the FP RSs is increased compared to the linear case as the reflection is further reduced at the surface due to the smooth permittivity. There are still high quality WG modes, with a decreased splitting between TE and TM RSs. Looking at the potential \Fig{f:modes_and_potential}f, we find the well further inside the sphere with a wide barrier extended towards the outside, which provides good containment for the RSs.  Higher-order TM WG modes along with the corresponding TM effective potentials are shown in \Fig{f:new_potential_and_excited_states} of \App{a:new_potential}.

\section{TE-TM splitting}
\label{s:tetm_degeneracy}

\begin{figure}[t]
	\centering
	\includegraphics[width=1\linewidth]{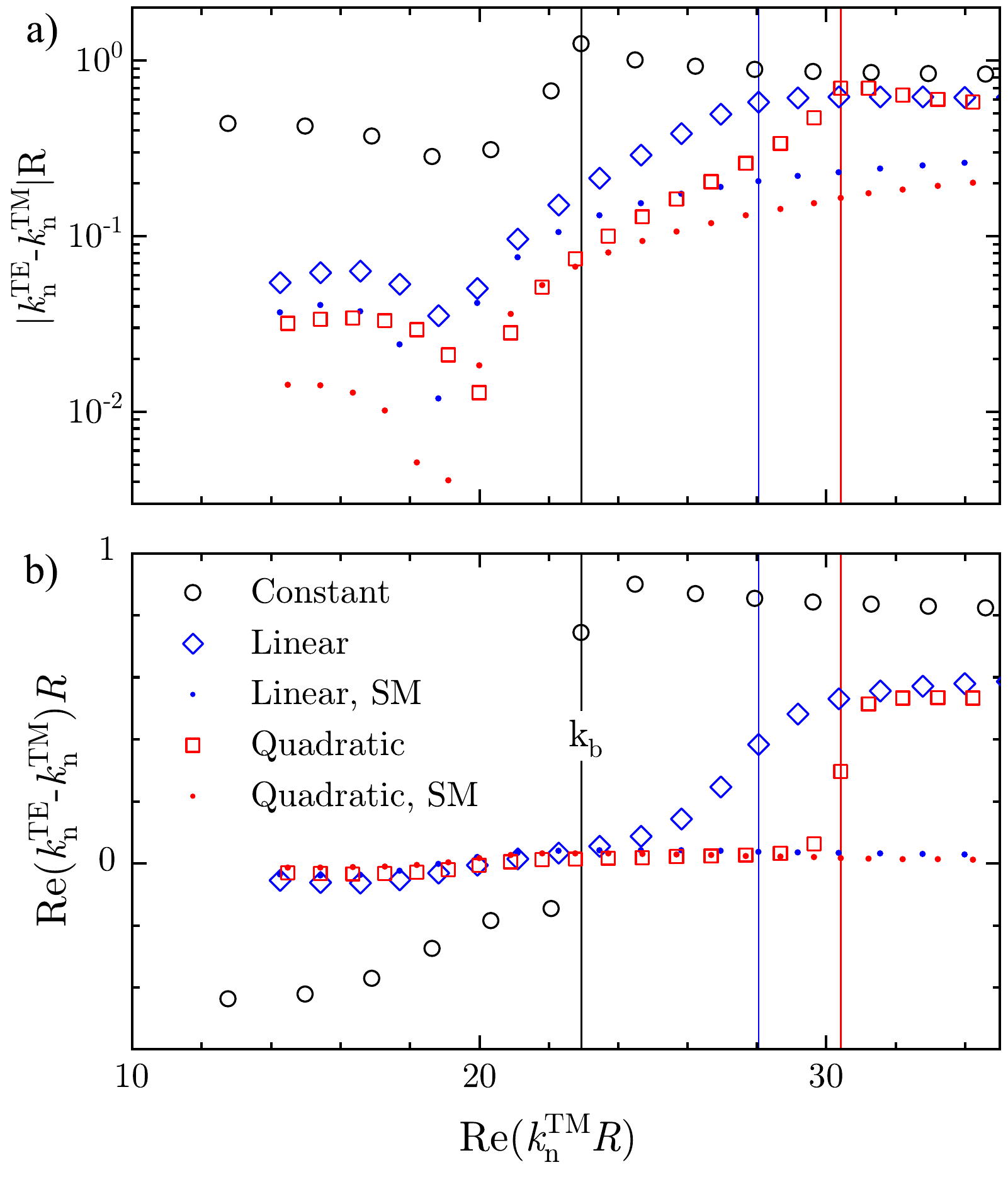}
	\caption{ Absolute value (a) and the real part (b) of the splitting between a TE ($k_{\rm TE}$) and the nearest TM ($k_{\rm TM}$) RS, for the considered permittivity profiles and $l=20$. The vertical lines are the positions of the Brewster peak ($k_b$) in each TM spectra. The single-mode (SM) values are based on
the re-expansion \Eq{eq:sm_reexpansion}.
}
\label{f:tetm_difference}
\end{figure}

The degeneracy of TE and TM modes might be of particular interest for chirality sensing, as that can convert second order perturbation effects due to a chiral material in the surrounding into the first order, similar to the effect of Faraday rotation by a circular magnetic field \cite{LanIEEE11, VincentAPL20}. 
We found in the previous section that for the linear and quadratic permittivity, the splitting between TE and TM RSs is reduced compared to the constant permittivity. This is quantified in \Fig{f:tetm_difference}, showing the distance from each TE RS to its nearest TM RS, both in the complex plane (\Fig{f:tetm_difference}a) and for the real part only (\Fig{f:tetm_difference}b). Considering first the constant permittivity, we find that the TE-TM splitting of WG modes is smaller than that of FP modes, and the real part of the splitting changes its sign at the Brewster peak, due to the additional TM mode as discussed in \Sec{ss:brewster}. At this peak there is a maximum of the absolute difference, due to the much larger imaginary part of the TM mode. 

Moving to the linear profile, the splitting decreases by a factor of about 5 for the WG modes, but only by about 30\% for FP modes. Consistent with the weaker Brewster peak in the spectrum (see \Fig{f:modes_and_potential}c), the splitting also does not show a pronounced peak. Finally, for the quadratic profile, the splitting is further reduced by a factor of about two for the WG modes and by about 10\% for the FP modes. Due to the larger imaginary part (see \Fig{f:modes_and_potential}e), also the absolute difference shows a Brewster peak. For all three cases, the smallest absolute distance between RSs is found for the WG modes near the critical wavenumber $k_c=l/R$ of the total internal reflection.

A similar behavior is observable for higher angular momentum numbers, as shown in \App{a:tetm_degeneracy_l80}. For higher $l$, it is also easer to see that the graded permittivity profile reduces the dispersion of the WG modes, creating an approximately equidistant spectrum as shown in \App{a:dispersion}. This has been also discussed in literature \cite{IlchenkoJOSAA03} and is consistent with results from the Morse potential approximation given by \Eq{eq:knTE_approximate}.

The RS splitting can be understood more mathematically by looking at the additional term of the TM potential in \Eq{eq:TM_potential}, which is the product of the logarithmic derivatives of the permittivity and the field. An obvious way to reduce the influence of this term is to spatially separate the maxima of the logarithmic derivative of the permittivity and the field amplitude. For the constant permittivity, the derivative creates a $\delta$ function at the boundary which overlaps much with the field thus creating a rather large splitting. Moving to the linear profile, the field maximum is shifted to smaller radii but the derivative of the permittivity is constant everywhere within the sphere. Still its influence is more spatially distributed compared to the $\delta$ function, and this reduces the splitting. Finally, for the quadratic profile, the maxima of both functions are spatially separated,  and this reduces  the splitting even further.

In the following subsection we quantify the influence of the additional term in the TM potential on a more rigorous level. A qualitative discussion of the TE-TM splitting of the fundamental WG mode
is provided in \App{a:tetm_degeneracy_l80}, in terms of the radial and polar confinement of light in an effective waveguide with an asymmetric cross-section.

\subsection{Perturbation from TE to TM}
\label{ss:perturbation_method}

The RSs form, together with static modes or their equivalents, a complete set inside the system and therefore provide a suitable basis for expanding any vector field within the system. This is the core principle of the RSE. In fact, an expansion into known basis modes is used in this paper to find the modes of the graded index profiles. In this subsection, we apply the same principle, however, in a simpler situation. Namely, we solve the scalar wave equation (\ref{eq:radial_equation_TM}) with the TM potential by expanding its solution into the complete set of eigenstates of the corresponding wave equation (\ref{eq:radial_equation_TE}) for TE polarization. In the simplest case, we reduce our basis to a single TE mode and thus solve \Eq{eq:radial_equation_TM} in the so-called diagonal approximation which can further be reduced to and interpreted as a first-order perturbation theory result.

The scalar equation (\ref{eq:radial_equation_TE}) for the TE RSs can be written as
\begin{equation}
\hat{L}(k_n,r) \mathcal{E}_n(r) = 0\,,
\end{equation}
where
\begin{equation}
\hat{L}(k,r) = \dv[2]{}{r} - \frac{\alpha^2}{r^2} + k^2 \varepsilon(r)\,.
\end{equation}
The corresponding scalar Green's function satisfies
\begin{equation}
\label{eq:greens_dyadic_reexpansion}
\hat{L}(k,r) G_k(r,r') = k \delta(r-r')
\end{equation}
and can be expanded as
\begin{equation}
\label{eq:gf_expansion}
G_k(r,r') = \sum_n \frac{\mathcal{E}_n(r) \mathcal{E}_n(r')}{k - k_n} = k \sum_n \frac{\mathcal{E}_n(r) \mathcal{E}_n(r')}{k_n(k - k_n)}\,,
\end{equation}
where $\mathcal{E}_n$ is normalized according to \Eq{norm-TE}, the same way as in \Onlinecite{MuljarovPRA20}. Accordingly, \Eq{eq:radial_equation_TM} for TM polarization takes the form
\begin{equation}
\hat{L}(k,r) \mathcal{H}(r) =  \Delta\hat{ L}(r) \mathcal{H}(r)\,,
\end{equation}
where
\begin{equation}
\Delta \hat{L}(r) = \frac{\varepsilon'(r)}{\varepsilon(r)} \dv{}{r}\,,
\end{equation}
and can be further written as a Lippmann-Schwinger equation, in terms the Green's function of the operator $\hat{L}(k,r)$:
\begin{equation}
\mathcal{H}(r) = \frac{1}{k}  \int_0^R G_k(r,r') \Delta \hat{L}(r') \mathcal{H}(r') \dd r'\,.
\label{eq:LS}
\end{equation}
Now, using the completeness of the basis states $\mathcal{E}_n(r)$,
\begin{equation}
\label{eq:H_expansion}
\mathcal{H}(r) = \sum_n c_n \mathcal{E}_n(r)\,,
\end{equation}
and the Green's function expansion \Eq{eq:gf_expansion}, we convert \Eq{eq:LS} into the following matrix equation
\begin{equation}
\label{eq:reexpansion}
k_n(k-k_n)c_n = \sum_{n'} \Delta_{nn'} c_{n'}\,,
\end{equation}
where
\begin{equation}
\Delta_{nn'} = \int_0^R \mathcal{E}_n(r) \frac{\varepsilon'}{\varepsilon} \mathcal{E}_{n'}'(r) \dd r
\end{equation}
and the primes in $\varepsilon$ and $\mathcal{E}$ mean derivatives with respect to $r$. Finally, using a single state only ($n'=n$), this reduces to the diagonal approximation:
\begin{equation}
\label{eq:sm_reexpansion}
k \approx k_n + \frac{\Delta_{nn}}{k_n}\,,
\end{equation}
which is clearly equivalent to the first-order result in terms of the perturbation matrix $\Delta_{nn'}$. We call the above method re-expansion as the basis functions $\mathcal{E}_n(r)$ used in the expansion \Eq{eq:H_expansion} are in turn expanded into the RSs of the homogeneous sphere.

A less rigorous and perhaps simpler approach is to treat the extra term in the TM potential, added to the TE equation, in a single mode approximation, in a manner it is usually applied to closed systems. Assuming $\mathcal{H}(r) \approx \mathcal{E}(r)$ and taking the difference between \Eqs{eq:radial_equation_TE}{eq:radial_equation_TM}, we find
\begin{equation}
	\left[- \frac{\varepsilon'(r)}{\varepsilon(r)} \dv[]{}{r} +(\kTM^2-\kTE^2) \varepsilon(r)  \right] \mathcal{E}(r) \approx  0\,,
\label{eq:tetm}
\end{equation}
where $\kTE$ ($\kTM$) is the TE (TM) RS wavenumber.
Multiplying \Eq{eq:tetm} with $\mathcal{E}(r)$ and integrating over the system volume yields
\begin{equation}
    \kTM^2 - \kTE^2 \approx \frac{\int_0^R \mathcal{E}(r)  \frac{\varepsilon'(r)}{\varepsilon(r)} \mathcal{E}'(r) \dd r }{\int_0^R \mathcal{E}(r) \varepsilon(r) \mathcal{E}(r) \dd r} \equiv 2\Delta \,.
\label{eq:tetm2}
\end{equation}
The first-order correction to the wavenumber, determining the TE-TM splitting is then given by
\begin{equation}
\label{eq:approximate_difference}
    \kTM \approx \kTE +\frac{\Delta}{\kTE}\,.
\end{equation}
For high-quality WG modes, the field $\mathcal{E}(r)$ is small at the surface, so that the integral in the denominator of \Eq{eq:tetm2} is getting close to the exact normalization,  $2 \int_0^R \varepsilon(r) \mathcal{E}^2(r) \dd r \approx 1$, and the two results, \Eqs{eq:sm_reexpansion}{eq:approximate_difference}, become identical.

We evaluate the TE-TM mode splitting using the diagonal approximation  \Eq{eq:sm_reexpansion} for the linear and quadratic profiles and compare it with the accurate RSE result in \Fig{f:tetm_difference}. The obtained values from the single mode approximation are in qualitative but not quantitative agreement with the RSE result, and for the WG modes about a factor of two smaller. So interestingly, while the TE-TM splitting is small, suggesting that the single mode approximation should be suitable, the TE and TM field distributions are actually significantly different. This is due to a rather large perturbation of the potential (see \Fig{f:modes_and_potential}), showing both positive and negative regions, and thus mixing with other modes while having a small single-mode perturbation integral.

\subsection{Wide potential well}
\label{ss:wide_well}

\begin{figure}[!htp]
	\centering
	\includegraphics[width=\linewidth]{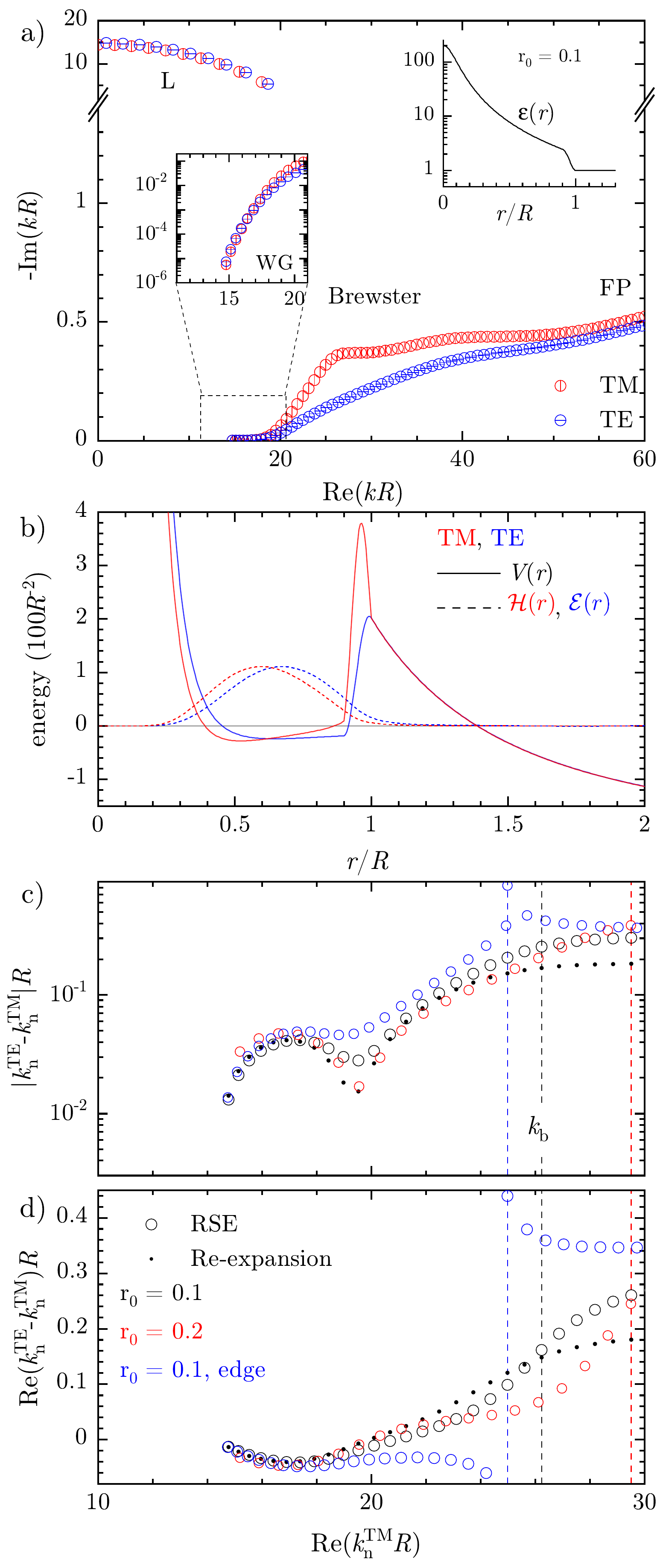}
	\caption{RSs in a graded index profile \Eq{eq:permittivity_wide} creating a wide potential well. a) as \Fig{f:modes_and_potential} left column. b) as \Fig{f:modes_and_potential} right column. c) and d) as \Fig{f:tetm_difference} but using
\Eq{eq:reexpansion} with $N=100$ basis modes, in comparison with the full RSE. As the imaginary part of $k_n$ is monotonously increasing from WG to FP modes, the Brewster peak value $k_b$ is chosen at the mode that has the largest difference of Im\,$k$ between the closest TE and TM modes.
}
\label{f:widewell}
\end{figure}

We expect the TE-TM degeneracy may be reduced for a wider potential well, as this can decrease the overlap of the RS field with the gradients of the permittivity, thus reducing the perturbation of the potential  treated in \Sec{ss:perturbation_method}. To create such a well in the effective potential $V^{\rm TE}$, given by \Eq{eq:TE_potential}, the centrifugal radial term $\alpha^2/r^2$ has to be compensated by a permittivity with the same functional dependence, $\varepsilon(r)\propto 1/r^2$. In this case the refractive index  $n(r)$ scales as $1/r$, so that the circular round-trip phase, $2\pi krn(r)$, which is equal to $2\pi l$ in the ray picture, is independent of $r$. In other words, this graded index creates equal optical ray path lengths at all radii. 

Since a permittivity diverging towards the sphere centre is not realistic, we introduce a cut-off radius $r_0 \ll R$ at which the permittivity saturates, using the expression
\begin{equation}
	\epsw(r) = \epsww \frac{R^2+r_0^2}{r^2+r_0^2}\,.
\end{equation}
Here $\epsww$ is the permittivity at the sphere surface $r=R$. In order to create a smooth potential with no discontinuities up to the first derivative across the sphere surface, we further introduce a transition region of width $r_0$ by defining the permittivity as
\begin{equation} \label{eq:permittivity_wide}
	\varepsilon(r) =
	\begin{cases}{}
	1 & r>R\,,\\
	\epsw(r)	& r < R-r_0\,, \\
	 1+[\epsw(r)-1]\sin^2\bigl(\pi\frac{r-R}{2r_0}\bigr)   &\mbox{otherwise\,.}
	 \end{cases}
\end{equation}
The resulting permittivity profile and RSs for $\epsww=2$ and $r_0=0.1R$ are shown in \Fig{f:widewell}a, calculated by the RSE with $N=1600$. The FP RSs are packed more densely than in the previous cases, due to the higher permittivity. The Brewster peak is blended in with the rest of the TM RSs, which have a monotonously increasing imaginary part; however we can still identify the peak in the difference of the imaginary part compared to the TE RSs. The potential for the first WG mode (\Fig{f:widewell}b) shows a wide and flat well, as designed. The splitting between TE and TM RSs (see black on \Fig{f:widewell}c) has reduced overall compared to the other profiles considered, and now the smallest absolute distance is observed for the first WG mode, being about twice smaller than for the quadratic profile (see \Fig{f:tetm_difference}). Increasing $r_0$ reduces the well width leading to larger splitting (see red on \Fig{f:widewell}c,d). Using a sharp boundary at the edge, i.e. without the $\sin^2$ term in \Eq{eq:permittivity_wide}, the splitting of the first mode is not significantly changed, as it has a small field at the boundary. Higher order modes instead acquire a larger splitting, and furthermore a sharper Brewster's peak is found (see blue on \Fig{f:widewell}c,d).

We also calculated the splitting using the perturbation method introduced in \Sec{ss:perturbation_method}. While the degeneracy in $k$ is decreased, the TE and TM fields are still spatially separated, so that instead of using a single mode we evaluate the full matrix equation \Eq{eq:reexpansion} for $N=100$ RSs. On \Fig{f:widewell} we can see that this leads to a much better agreement with the results compared to the single mode approximation used for the linear and quadratic case before. For increasing $k$ the error in the results increases. This is due to a combination of factors, including the truncation of the matrix, the slow convergence of the expansion \Eq{eq:gf_expansion} as discussed in \Onlinecite{MuljarovPRA20}, and the error in the unperturbed fields $\mathcal{E}_n$.

\section{Summary}
\label{s:conclusion}

We have studied, for different static-mode sets, an optimized version of the resonant-state expansion (RSE) and demonstrated the same quick ($1/N^3$, where $N$ is the basis size of the RSE) convergence to the exact solution for different static-mode sets. We have also compared it with a similar version of the RSE, studied earlier in~\cite{MuljarovPRA20}, in which static modes are eliminated from the basis, and demonstrated the same convergence for both versions. We have then applied the RSE to spheres with graded permittivity profiles and shown that the RSE is a reliable and simple method to determine all the resonant states (RSs) up to a maximum wavenumber controlled by the basis choice. Looking at the full spectrum provided by the RSE, instead of just distinct RSs, allows us to identify physical phenomena reliably and rapidly, as shown by the results presented. We have further discussed the results using the ray picture with surface reflections, the phase analysis based on the secular equation, and the concept of an effective potential, treating the radial wave equation as a quantum-mechanical analogue. Importantly, we provide a MATLAB program to calculate modes of a spherically symmetric system with a polynomial permittivity profile. Once the basis modes are calculated across the whole system volume, applying the perturbation and finding the new modes takes only a few seconds on a modern computer, therefore the RSE is particularly suited to explore large parameter spaces.

For a homogeneous sphere, we have provided a detailed analysis of the spectrum of the RSs in the complex wavenumber plane, consisting of leaky, Fabry-P\'erot (FP), and whispering-gallery (WG) modes. This analysis includes development of a number of approximations. For the transverse-magnetic (TM) polarization, we have explained the peak in the RS linewidth and an additional mode with respect to the transverse-electric (TE) polarization in terms of the Brewster phenomena. Using the ray picture further, we have evaluated the RS linewidth from Fresnel's coefficients of reflection which provides a good agreement with the exact solution for FP modes. We have shown that the wavenumber $k_c=l/R$ evaluated at the critical angle of the total internal reflection plays the role of a boundary in the spectrum separating the WG from FP modes. Using the phase analysis of the secular equation, we developed an analytic approximation for the WG and FP mode linewidths, an asymptotic formula for the FP wavenumbers at large frequencies, and have shed light on the mode separation and TE-TM splitting.

We have then investigated graded index spheres with linear or quadratic permittivity profiles eliminating the discontinuity at the sphere surface. We have found that the imaginary part of FP modes is increasing logarithmically with their wavenumber, with a larger slope for quadratic profiles. 
We have used the concept of effective potential for the radial electro-magnetic wave equation and suggested an interpretation of this quantum-mechanical analogy by associating all the physical solutions with zero-energy states, emphasizing that the effective potentials are complex. This provides a clear qualitative picture explaining the existence and properties of WG modes. 
We have further approximated the obtained effective potentials around their minimum with the analytically solvable Morse potential, which for TE polarization yields a simple explicit algebraic expression of high accuracy for the WG mode wavenumbers. For large angular quantum numbers $l$, this solution predicts a nearly equidistant spectrum of WG modes, similar to that of a harmonic oscillator.

We have studied the TE-TM splitting and demonstrated its reduction for WG modes when going from constant to linear and then to quadratic permittivity profile. We have shown that the splitting is further reduced in a wide flat potential well designed via the radial permittivity. To understand the TE-TM splitting, we have developed a re-expansion method, which perturbatively treats the difference between the effective potentials of TE and TM polarizations. The results are in good agreement with the exact solution. We have also provided a diagonal approximation, which turns out to be insufficient for the investigated cases despite the small splitting -- a consequence of the underlying strong perturbation.

\begin{acknowledgments}
 Z.S. acknowledges the Engineering and Physical Sciences Research Council for his PhD studentship award (grant EP/R513003/1).
\end{acknowledgments}

\appendix

\section{Phase analysis for a sphere}
\label{a:phase_analysis}

\begin{figure}
	\includegraphics[width=1\linewidth]{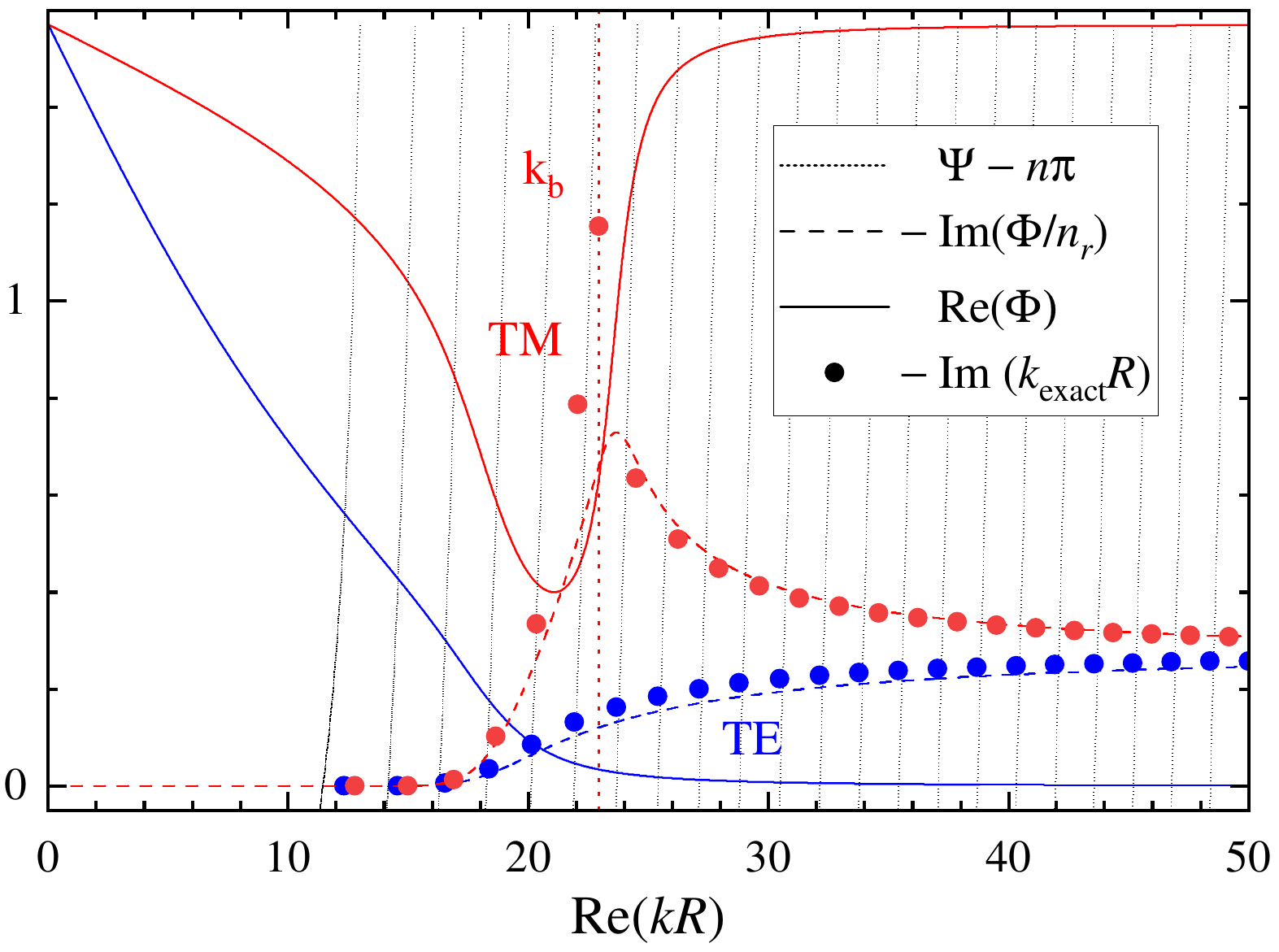}
	\caption{Phase functions $\Psi({\rm Re}\,k)$ (black dotted lines), Re$(\Phi({\rm Re}\,k))$ (solid lines) and Im\,$(\Phi({\rm Re}\,k))/n_r$ (dashed lines) for TE and TM RSs, alongside the exact RS wavenumbers (blue and red dots) in the complex wavenumber plane, for a homogeneous sphere of $n_r=2$ and $l=20$. The vertical dotted line shows the Brewster wavenumber $k_b$.
}
	\label{f:phase}
\end{figure}

\begin{figure}[!t]
	\includegraphics[width=1\linewidth]{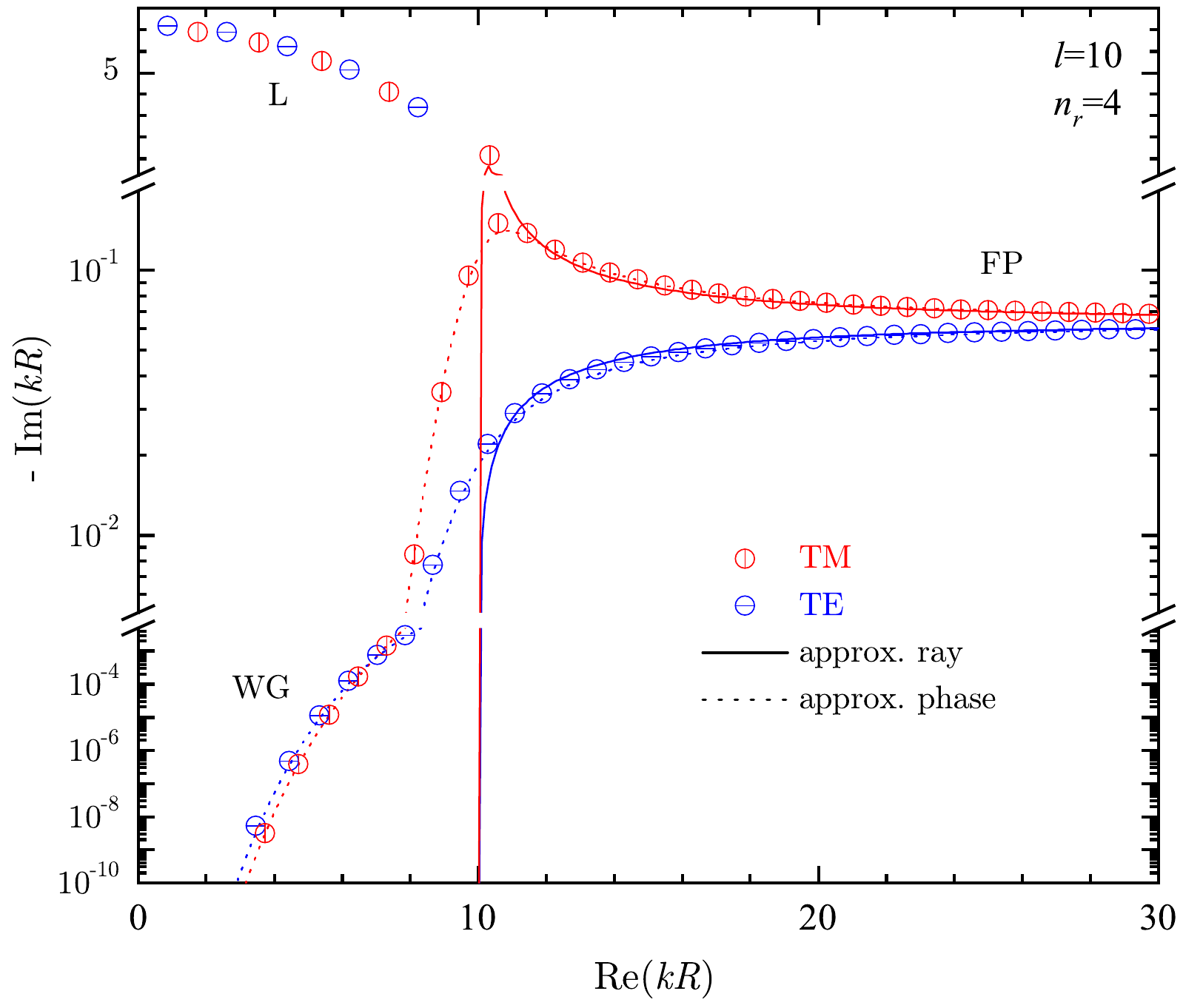}
	\caption{As \Fig{f:sphere} but for $n_r=4$ and $l=10$.}
\label{f:sphere_app}
\end{figure}

The RS wavenumbers of a homogeneous sphere in vacuum are determined by the secular equation (\ref{eq:secular_equation}). Its approximate solution \Eqs{eq:phipsi}{eq:kpp} developed in \Sec{ss:phase_analysis} is illustrated in \Fig{f:phase}. The black dotted lines show
$\Psi({\rm Re}\,k)-n\pi$, for all values of $n$, while blue and red solid lines show the real part of $\Phi({\rm Re}\,k)$ for TE and TM polarizations, respectively. According to \Eq{eq:phipsi}, they should cross the black dotted lines at the real part of the RS wavenumbers, Re\,$k_n$, whereas the imaginary part  Im\,$k_n$ is approximately given by Im$(\Phi({\rm Re}\,k_n))/n_r$ (blue and red dashed lines), according to \Eq{eq:kpp}. Generally, it can be seen a good agreement with the exact values shown by blue and red dots, representing the RS wavenumbers in the complex $k$-plane. At large $kR$, Re\,$\Phi(k)$ approaches the asymptote at 0 ($\pi$) for TE (TM) polarization, which determines the mode separation, in accordance with \Eq{eq:mode_approximate}. For a twice larger refractive index of the sphere ($n_r=4$), and $l$ reduced to 10 in order to create a similar number of WG modes, an improved agreement between this approximation and the exact solution is found, as shown in \Fig{f:sphere_app}.

To derive the large-$k$ approximation given by \Eq{eq:mode_approximate}, we first note that
for $z\gg l$,
\begin{equation}
\frac{H'(z)}{H(z)} \approx i\,.
\end{equation}
Introducing $\tilde z = n_r z - (l+1)\pi/2$, where $z=kR$,
and also using the approximation \Eq{eq:tan_approximation}, the secular equation (\ref{eq:secular_equation}) takes the form
\begin{equation}
\label{a:eq:approx}
\tan(\tilde z) \approx -\frac{i}{\beta}\,,
\end{equation}
which can be also written as
\begin{equation}
e^{2i\tilde z} \approx \frac{1 + 1/\beta}{1- 1/\beta}\,.
\label{a:eq:approx_mod}
\end{equation}
This equation has explicit analytical solutions
\begin{align}
\label{a:eq:mode_approx}
\tilde z^{\rm TE}_n &\approx \pi n - \frac{i}{2} \ln{\frac{n_r + 1}{n_r -1}}\,, \nonumber\\
\tilde z^{\rm TM}_n &\approx \pi \left(n + \frac{1}{2}\right) - \frac{i}{2} \ln{\frac{n_r + 1}{n_r -1}}\,,
\end{align}
equivalent to \Eq{eq:mode_approximate}. The TE result was also given in \Onlinecite{BraunsteinSPIE96}. Note that apart from the \mbox{$-(l+1)\pi/2$} term in $\tilde z$, these are the same as the modes of a homogeneous slab at normal incidence~\cite{MuljarovEPL10}. The TE (TM) modes correspond to the odd (even) modes of the slab, as discussed in more depth in \App{a:homogeneous_slab} below. From here we find in particular that the wavenumber difference between neighboring modes in a given polarization is $\pi / n_r R$, consistent with the graphical solution in \Fig{f:phase}. We can also see that the difference between neighboring TE and TM FP RSs is
\begin{equation}
\label{a:eq:mode_spacing}
\Delta \tilde z = n_r (k^{\rm TM} - k^{\rm TE}) R = \frac{\pi}{2}\,,
\end{equation}
as also suggested by \Fig{f:phase}.

In principle, a similar result can be derived for WG modes in the case when $n_r \gg 1$. The latter condition allows the argument of the Bessel functions ($n_rkR$) to be large (compared to $l$), leading to the approximation \Eq{eq:tan_approximation}, while simultaneously keeping the argument of the Hankel function small (compared to $l$). In this case $H'(z)/H(z) \approx -l/z$, which in the WG limit gives a modified equation compared to \Eq{a:eq:approx}:
\begin{equation}
\label{a:eq:approx2}
\tan(\tilde z) \approx \frac{l}{\beta z}\,.
\end{equation}
Therefore it is possible to observe in a very high permittivity material nearly equidistant WG modes even in a homogeneous sphere. This is consistent with \cite{RollJOSAA00}, where the resonances positions and mode separations were described based on geometrical optics, and also with approximate results from \cite{ProbertJonesJOSAA84} for the mode spacing when $l \gg 1$.

\section{Eigenmodes of a homogeneous slab}
\label{a:homogeneous_slab}

By approximating the surface of the sphere with a flat boundary, we compare the modes of a sphere with those of a homogeneous slab, in which EM waves propagate at a non-normal incidence to the boundary.
We also compare here the modes of the slab with an approximation similar to \Eq{eq:imk} which is provided by the ray picture.

The secular equation determining the TE modes of a homogeneous slab of thickness $2a$, permittivity $\epsilon$, and permeability $\mu$ is given by~\cite{NealePRB20}
\be
\label{eq:homogeneous_slab}
 e^{2iq_n a} = (-1)^n \frac{q_n + \mu k_n }{q_n - \mu k_n}\,,
\ee
where $q=\sqrt{\epsilon \mu \omega^2/c^2- p^2}$ and $k=\sqrt{\omega^2/c^2 - p^2}$ are the normal components of wavenumber inside the slab and in vacuum,  respectively, and $p$ is its in-plane component,  which is conserved, so that $p$ is essentially the same as the one sketched in \Fig{f:brewster}. The factor $(-1)^n$ gives the mode parity and can be used to label the modes. The corresponding equation for TM modes is provided by just swapping $\epsilon$ and $\mu$ in \Eq{eq:homogeneous_slab}.
The similarity between \Eqs{eq:homogeneous_slab}{a:eq:approx_mod} is obvious. Clearly, these equations become identical for normal incidence, when $p=0$ and consequently $q=n_r k$ with $n_r=\sqrt{\epsilon \mu}$.

One can find an approximate imaginary part of the mode wavenumbers in the same way as described in \Sec{ss:brewster}. The angle of incidence inside the slab is given by $\theta = \atan(p/q)$, and the optical path length is $L = 2an_r/\cos\theta$ where $a$ is the slab half width. The imaginary part of the RS wavenumbers is then given by
\begin{equation}
\label{eq:approximate_imaginary}
\Im k = \frac{\ln|r_P|}{2an_r}\cos\theta   \,,
\end{equation}
where again $r_P$ is the polarization dependent Fresnel coefficient taken at real wavenumbers -- compare  \Eqs{eq:approximate_imaginary}{eq:imk}.

\begin{figure}[!t]
	\includegraphics[width=1\linewidth]{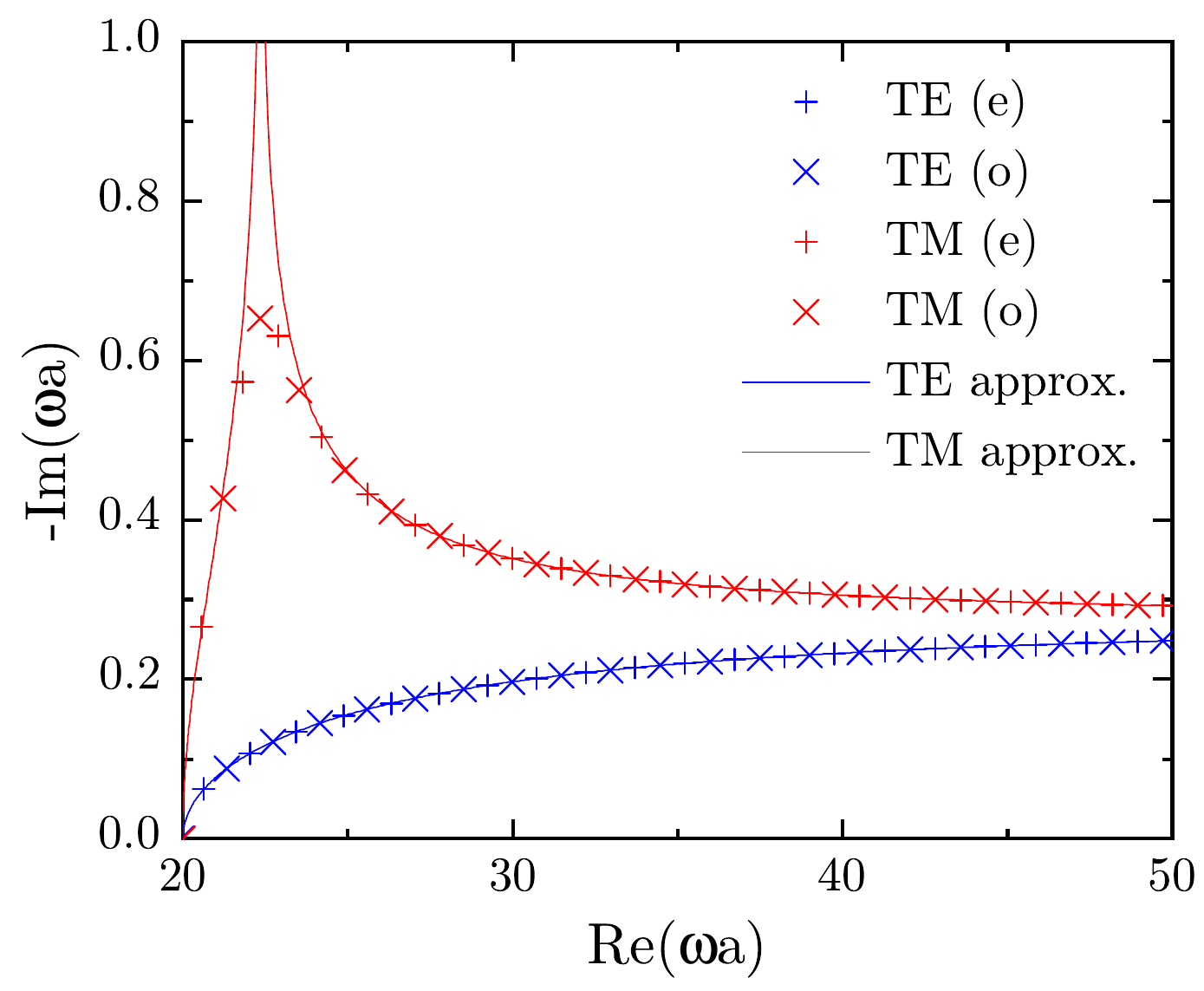}
	\caption{Eigenmodes of a homogeneous slab with $\epsilon=4$, $\mu=1$, and $p=20$, along with approximate solutions for the imaginary part obtained from \Eq{eq:approximate_imaginary}. The `(e)' and `(o)' label the even and odd modes, respectively.}
\label{f:homogeneous_slab}
\end{figure}

We show in \Fig{f:homogeneous_slab} the TE and TM modes of a slab with permittivity $\epsilon=4$, permeability $\mu=1$, and in-plane wavenumber $p=20/a$, so that the system parameters are matching those used for the sphere in \Sec{ss:brewster}. In the TM spectrum, there is a peak again, which is aligned with the position of the Brewster angle. Overall, for these parameters the approximation works better for the slab than for the sphere, as the boundary is strictly flat in this case. The observed small deviation of the modes from the approximation at the Brewster peak is due to that the imaginary part of the wavenumber is neglected in $r_P$, which can be significant around the peak. We can see that for both polarizations, there are even and odd modes following each other in alternating order. We can also see that at high frequencies, both TE and TM FP modes converge to the same asymptote, which is the same behavior as for the sphere. As in the spherical case, the TE and TM modes of the same parity appear in alternating order. In fact, even TE modes align with the odd TM modes, and vice versa.
However, in case of a sphere even modes do not exist. Finally, instead of the WG modes of a sphere, in case of a slab there are waveguide modes with purely real eigenfrequencies, formed as a consequence of total internal reflection at the planar boundary, and instead of the L modes of a sphere, there are anti-waveguide modes in case of a slab~\cite{ArmitagePRA14}. Both waveguide and anti-waveguide modes have $\omega < pc$ and are not shown in \Fig{f:homogeneous_slab}.

\section{Resonant-state expansion for spherically symmetric systems}
\label{a:RSE}

According to \Onlinecite{MuljarovPRA20}, the matrix equation of the RSE for non-dispersive systems has the following general form:
\be
(k-k_n)a_n=-k\sum_{n'}\tilde{V}_{nn'} a_{n'}\,,
\label{RSE-gen}
\ee
where $a_n$ are the expansion coefficients of a perturbed RS into the basis RSs labeled by index $n$. For spherically symmetric systems, all $n$ refer to the same spherical quantum numbers $l$ and $m$, but the matrix elements $\tilde{V}_{nn'}$ of the perturbation are quite different in TE and TM polarizations.

For a radially-dependent permittivity perturbation $\Delta\varepsilon(r)$ of a nonmagnetic system,
the matrix elements in TE polarization are given by
\be
\tilde{V}_{nn'}^{\rm TE}=\int_0^R\cE_n(r)\Delta\varepsilon(r)\cE_{n'}(r)\dd r\,,
\label{me-TE}
\ee
where $\cE_n(r)$ is the electric field of the basis RS $n$, satisfying \Eq{eq:radial_equation_TE}, in which $k=k_n$ is the RS wavenumber and $\varepsilon(r)$ is the permittivity profile of the basis system. The fields $\cE_n(r)$  are normalized according to~\cite{MuljarovEPL10,MuljarovPRA20}
\be
2\int_0^R \varepsilon \cE_n^2\dd r+\frac{1}{k_n}\left[\left(\cE_n r\cE_n'\right)'-2r(\cE_n')^2\right]_{r=R}
=1\,.
\label{norm-TE}
\ee

For TM polarization, the matrix elements have a more complex form:
\be
\tilde{V}_{nn'}^{\rm TM}={V}_{nn'}-\sum_{jj'} V_{nj} W_{jj'} V_{j'n'}
\label{me-TM}
\ee
where $W_{jj'}$ is the inverse of matrix $\delta_{jj'}+ V_{jj'}$, index $n$ labels the basis TM RSs, and index $j$ labels additional functions required for completeness. They are used in the expansion of the perturbed EM vector fields and the dyadic GF, and are responsible for the static pole representation of the latter~\cite{MuljarovPRA20}. It is convenient to introduce a combined index $\nu$ which labels together the RSs ($n$) and the additional basis functions  ($j$). It is also useful to separate each basis electric vector field into the radial $\cE_\nu^r(r)$ and tangent $\cE_\nu^t(r)$ components. The matrix elements contributing to \Eq{me-TM} then take the form~\cite{MuljarovPRA20}:
\be
V_{\nu\nu'}=\int_0^R\left[\cE^t_\nu\Delta\varepsilon(r)\cE^t_{\nu'}+
\cE^r_\nu\frac{\varepsilon(r)\Delta\varepsilon(r)}{\varepsilon(r)+\Delta\varepsilon(r)}\cE^r_{\nu'}\right]\dd r
\label{me-TM2}
\ee
with $\cE_\nu^t(r)$ and $\cE_\nu^r(r)$ defined below.

For the basis TM RSs, the fields are given by
\be
\left(\begin{array}{cc}
\cE_n^t(r)\\
\cE_n^r(r)
\end{array}\right)
= -\frac{1}{k_n \varepsilon(r)}
\left(\begin{array}{cc}
\dv{}{r}\\
\frac{\alpha}{r}
\end{array}\right)
\cH_n(r)\equiv
\left(\begin{array}{cc}
\cK_n(r)\\
\cN_n(r)
\end{array}\right)
\,,
\label{modes-TM}
\ee
where $\cH_n(r)$ is the magnetic field of the basis TM RS $n$, satisfying \Eq{eq:radial_equation_TM}, in which $k=k_n$ is the RS wavenumber and $\varepsilon(r)$ is the permittivity profile of the basis system. The fields $\cH_n(r)$  are normalized according to~\cite{MuljarovEPL10,MuljarovPRA20}
\be
2\int_0^R \cH_n^2\dd r+\frac{1}{k_n}\left[\left(\cH_n\frac{r}{\varepsilon(r)}\cH_n'\right)'-\frac{2r}{\varepsilon(r)}(\cH_n')^2\right]_{r=R_+}
=1
\label{norm-TM}
\ee
with $R_+=R+0_+$, where $0_+$ is a positive infinitesimal.

All other basis states can be expressed in terms of functions $\cK_n(r)$ and $\cN_n(r)$ introduced in \Eq{modes-TM} and static modes $\psi_\lambda(r)$ introduced in~\cite{LobanovPRA19} and also discussed in \cite{MuljarovPRA20}. Let us note at this point that the two slightly different versions of the efficient (i.e. quickly convergent) RSE developed in \Onlinecite{MuljarovPRA20} are based on two different Mittag-Leffler representations of the full dyadic GF of a spherically symmetric system, called in \cite{MuljarovPRA20} ML3 and ML4. Essentially, they differ in the basis functions describing the static pole of the GF. Also, ML4 is introduced for a homogeneous sphere only, while ML3 is valid for any spherically symmetric basis system.

In the ML3 version of the RSE, all the additional basis states can be divided into three groups. In the first two groups, indices $j_{\rm I}$ and $j_{\rm II}$ take the same values as the TM RS index $n$, and the fields are given by
\be
\left(\begin{array}{cc}
\cE_{j_{\rm I}}^t\\
\cE_{j_{\rm I}}^r
\end{array}\right)
=
\left(\begin{array}{cc}
i\cK_n\\
i\cN_n
\end{array}\right)
\,\mbox{and}\,
\left(\begin{array}{cc}
\cE_{j_{\rm II}}^t\\
\cE_{j_{\rm II}}^r
\end{array}\right)
=
\left(\begin{array}{cc}
\cK_n\\
0
\end{array}\right)
\,.
\label{groups12}
\ee
In the third group,
\be
\left(\begin{array}{cc}
\cE_{j_{\rm III}}^t\\
\cE_{j_{\rm III}}^r
\end{array}\right)
=
\left(\begin{array}{cc}
\alpha\psi_\lambda\\
0
\end{array}\right)
\,,
\label{group3}
\ee
and the index $j_{\rm III}$ coincides with $\lambda$ labeling static modes defined in terms of the radial part of their potential function $\psi_\lambda(r)$.  Static modes are the solutions of a generalized Sturm-Liouville problem~\cite{LobanovPRA19,MuljarovPRA20} and are normalized according to
\be
\lambda^2\int_0^R \varepsilon(r)\psi^2 _\lambda(r)r^2 \dd r =1\,.
\label{norm-static}
\ee

For a basis system in the form of a non-magnetic homogeneous sphere in vacuum, described by the permittivity profile given by
\be
\varepsilon(r)=(\epsilon-1)\theta(R-r)+1\,,
\ee
the static mode potentials take the explicit form
\begin{equation}
\label{eq:static_potential}
\psi_\lambda(r) = A_\lambda j_l(\lambda r)
\end{equation}
 within the sphere ($r\leqslant R$), where $j_l(x) $ is the spherical Bessel function of order $l$,
$\lambda$ is the mode eigenvalue (here also used to label the modes), and $A_\lambda$ is a normalization constant determined according to \Eq{norm-static}. The eigenvalues $\lambda$ are found from the boundary condition of the Sturm-Liouville problem~\cite{LobanovPRA19}, which leaves a large range of possible sets. Following~\cite{MuljarovPRA20}, we consider here three sets of static modes for ML3 version of the RSE: (i) the volume-charge set (VC), with the eigenvalues generated by the secular equation
\begin{equation}
\lambda \epsilon R j_l'(\lambda R) + (l+1) j_l(\lambda R) = 0\,,
\end{equation}
(ii) the volume-surface-charge set (VSC), with a simpler secular equation
\begin{equation}
 j_l(\lambda R) = 0\,,
\end{equation}
and  (iii) a modified-volume-surface-charge set (MVSC), determined by the following secular equation
 \begin{equation}
\lambda R j_l'(\lambda R) + (\epsilon l+1) j_l(\lambda R) = 0.
 \end{equation}
Note that apart from the modes generated by the secular equations, both VSC and MVSC sets include one additional mode, that corresponds to $\lambda=0$. Also note that the VSC and VC sets were used in~\cite{LobanovPRA19} for a slowly convergent version of the RSE.

In the ML4 version of the RSE, developed in~\cite{MuljarovPRA20} for the basis system in a form of a homogeneous sphere in vacuum, all basis states responsible for the static pole of the GF can be divided into four groups. The first two groups are the same as in ML3 and are given by \Eq{groups12}. The third and fourth groups of basis functions provide an alternative to the static mode sets described above.
The third group is given by
\be
\left(\begin{array}{cc}
\cE_{j_{\rm III}}^t(r)\\
\cE_{j_{\rm III}}^r(r)
\end{array}\right)
=
\left(\begin{array}{cc}
\cN_n(r)\\
0
\end{array}\right)
\,,
\label{group3a}
\ee
where index $j_{\rm III}$ again takes the same values as the TM RS index $n$, in the same way as in the first two groups, and the fourth group consists of the single element
\be
\left(\begin{array}{cc}
\cE_{j_{\rm IV}}^t(r)\\
\cE_{j_{\rm IV}}^r(r)
\end{array}\right)
=
\left(\begin{array}{cc}
\cM_0(r)\\
0
\end{array}\right)
\,,
\label{group4}
\ee
where
\be
\cM_0(r)=\sqrt{\frac{l(l+1)}{\epsilon R} \frac{\epsilon-1}{\epsilon l+l+1}} \left(\frac{r}{R}\right)^l\,,
\ee
which can also be found as
\bea
\cM_0(r)&=&(\epsilon-1)\sqrt{\frac{l}{\epsilon}} \lim_{k_n\to0}
\cK_n(r)\\ &=&(\epsilon-1)\sqrt{\frac{l+1}{\epsilon}} \lim_{k_n\to0} \cN_n(r)\nonumber
\eea
by treating both $\cK_n(r)$ and $\cN_n(r)$ as analytic functions of $k_n$ and taking the limit $k_n\to0$.

\begin{figure}[!t]
	\centering
	\includegraphics[width=1\linewidth]{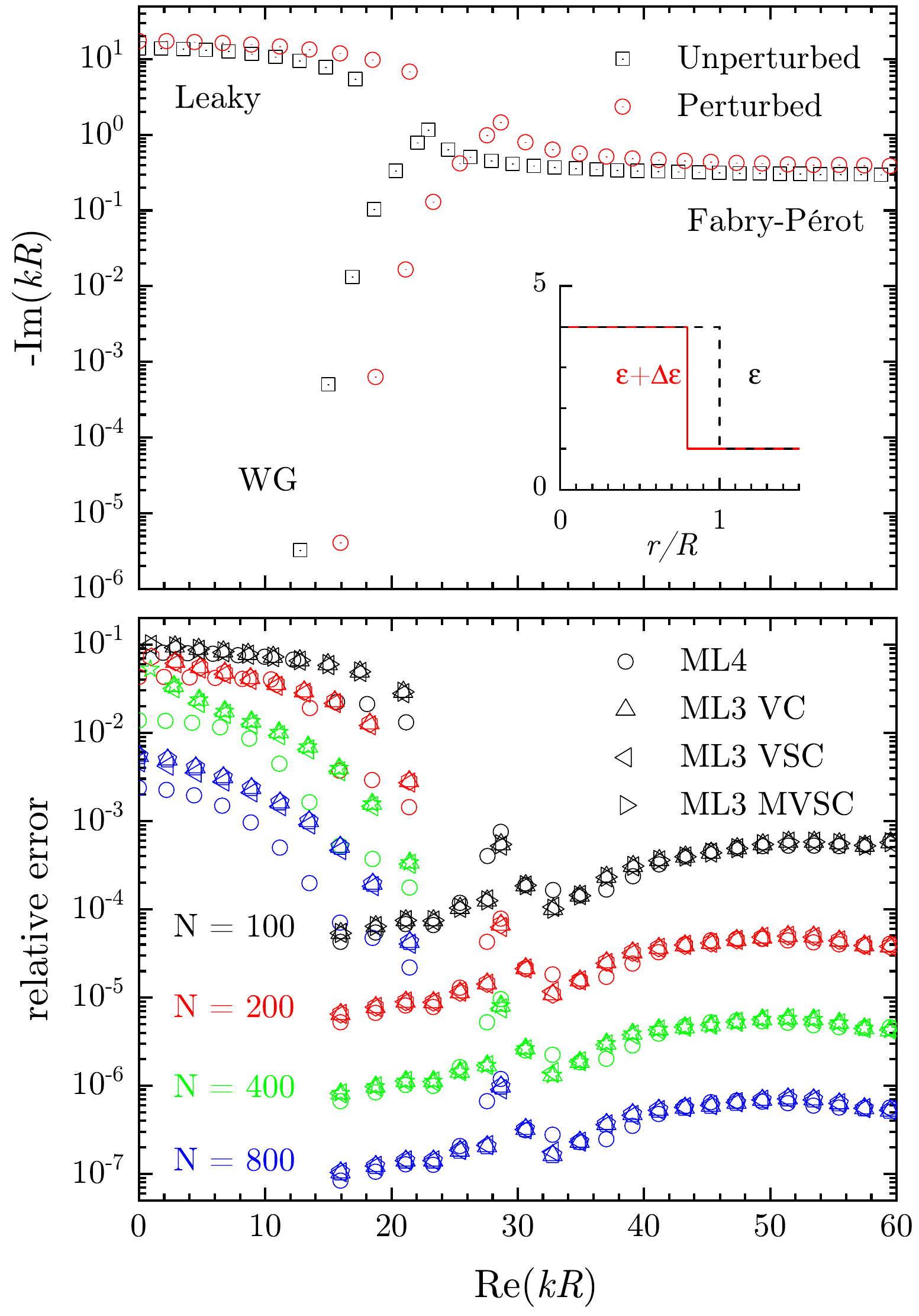}
	\caption{RSE applied for size perturbation. TM RSs of a dielectric sphere with permittivity $\epsilon = 4$ and permeability $\mu=1$, for angular momentum number $l = 20$. Top: Wavenumbers in the complex $k$ plane. Basis RSs for size $R$, target RSs for size $0.8R$. Bottom: Relative error of the RSs calculated by RSE with method of elimination of static modes (ML4) and with efficient inclusion of static modes (ML3) for various static mode sets and basis sizes $N$ as given. Inset: unperturbed and perturbed permittivity profiles.}
	\label{f:size_perturbation}
\end{figure}

To test the convergence of the RSE based on ML3 for the different static mode sets, we apply the RSE to a size perturbation of a homogeneous sphere. We choose as unperturbed system a homogeneous sphere in vacuum, having radius $R$, permittivity $\epsilon = 4$, and permeability $\mu=1$. We focus here on the TM RSs with angular momentum $l=20$, also noting that in spherically symmetric systems, all states are degenerate in $m$. The target system is a sphere of the same material and radius $0.8R$, so that the perturbation is given by $\Delta \varepsilon = 1 - \epsilon$ in the outer $0.2R$ thick shell of the basis sphere.
Figure~\ref{f:size_perturbation} shows the resulting perturbed and unperturbed eigenvalues $k$, and their error, for various basis sizes $N$, which include RSs with $|k_n|R \lesssim  0.77N$ and static modes with $|k_\lambda|R \lesssim 3.31N$. For a homogeneous sphere, in the absence of dispersion the RS wavenumbers $k_n$ and $R$ are inversely proportional, which can be seen as a scaling of the target RSs compared to the basis RSs in the complex plane.

The relative error for ML3 scales as $1/N^3$ (the same as in ML4), independent of which static mode set is used. In the original version of the RSE~\cite{LobanovPRA19}, with a slow ($1/N$) convergence for static mode inclusion, there was a more significant difference between the VC and VSC sets, as they were used for the expansion of the complete residue of the static pole of the GF, including the $\delta$-function term. We find that ML4 provides smaller errors for the leaky branch. This can be understood by noting that ML4 uses instead of static modes basis functions proportional to the RSs, including L RSs, and thus can be expected to be better suited for expanding the L RSs of the target system. A slow initial convergence of L RSs is testament to their unusual spatial shape, not well described by the basis RSs, but the $1/N^3$ convergence is eventually recovered above $N=400$.

\begin{figure}[!t]
	\centering
	\includegraphics[width=1\linewidth]{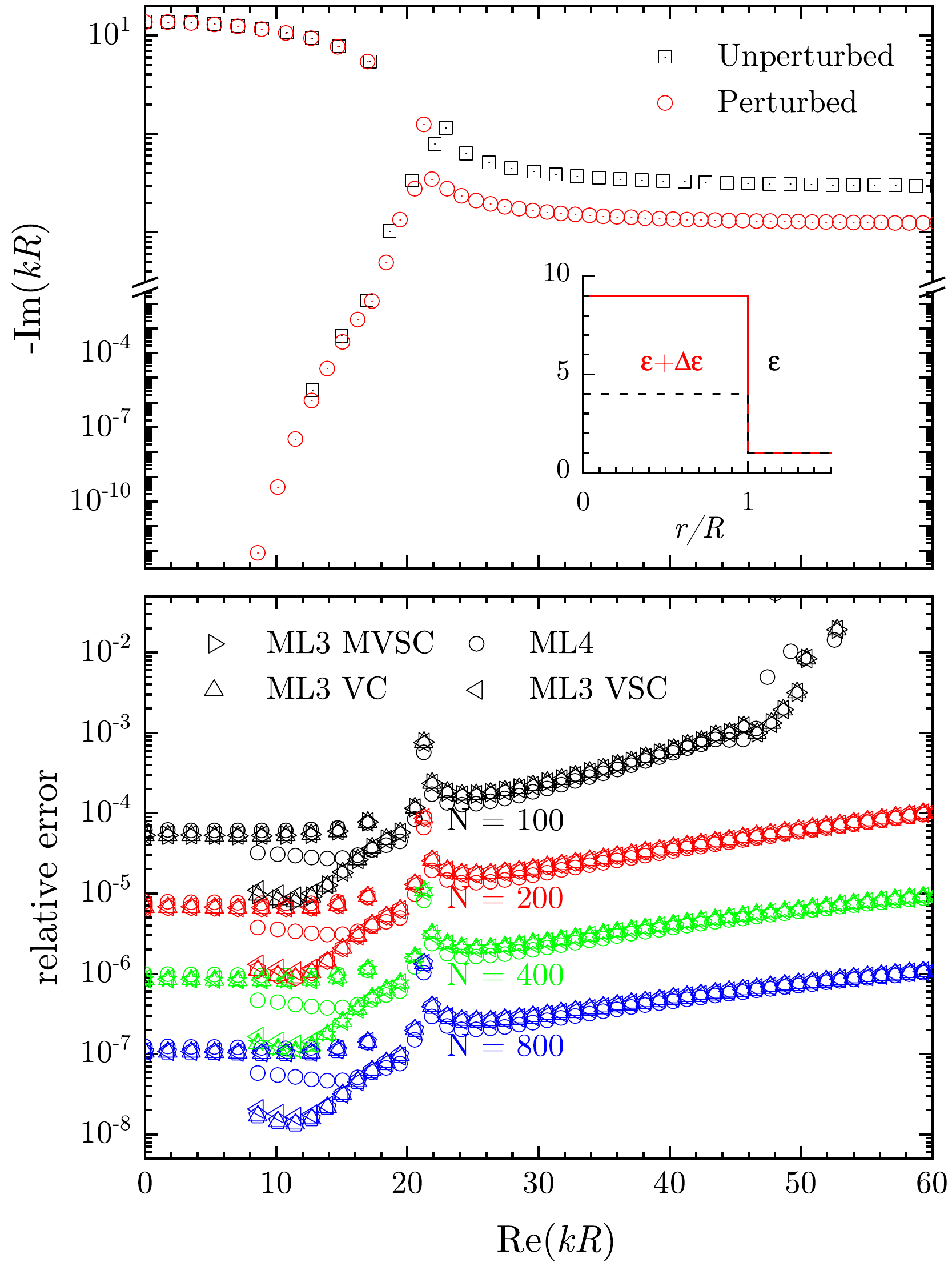}
	\caption{As \Fig{f:size_perturbation}, but for a homogeneous perturbation. Basis system $\epsilon = 4$, target system $\epsilon + \Delta \varepsilon = 9$.}
\label{f:strength_perturbation}
\end{figure}

The results for strength perturbation, that is, changing the permittivity of the sphere homogeneously, are shown in \Fig{f:strength_perturbation}, displaying a similar behavior. Here, using the same basis sizes as in \Fig{f:size_perturbation}, we apply the RSE for a homogeneous increase of the permittivity of the sphere by $\Delta \varepsilon = 5$, giving a target sphere permittivity $\epsilon + \Delta \varepsilon = 9$.  The higher refractive index leads to a denser array of RSs, increased number of WG modes and smaller imaginary part for the FP modes. We can see that the error converges with the basis size $N$ as $1/N^3$ for ML3, independent of the static mode set used. For the WG modes, the ML3 representation has some advantage over ML4, having up to five times smaller errors. The static modes thus seem better suited to describe these WG modes, likely because they are bound to the sphere, similar to the WG modes.

We thus conclude that for all three static mode sets, ML3 has a convergence similar to ML4. We used the ML4 version of the RSE for generating the results of this paper.

\section{Effective potential for TM modes}
\label{a:new_potential}

Here we show that the wave equation (\ref{eq:radial_equation_TM}) for the scalar magnetic field $\mathcal{H}(r)$ in TM polarization can be brought to a Schr\"odinger-like equation with an effective potential independent of the wave function.

\begin{figure}[!t]
	\centering
	\includegraphics[width=1\linewidth]{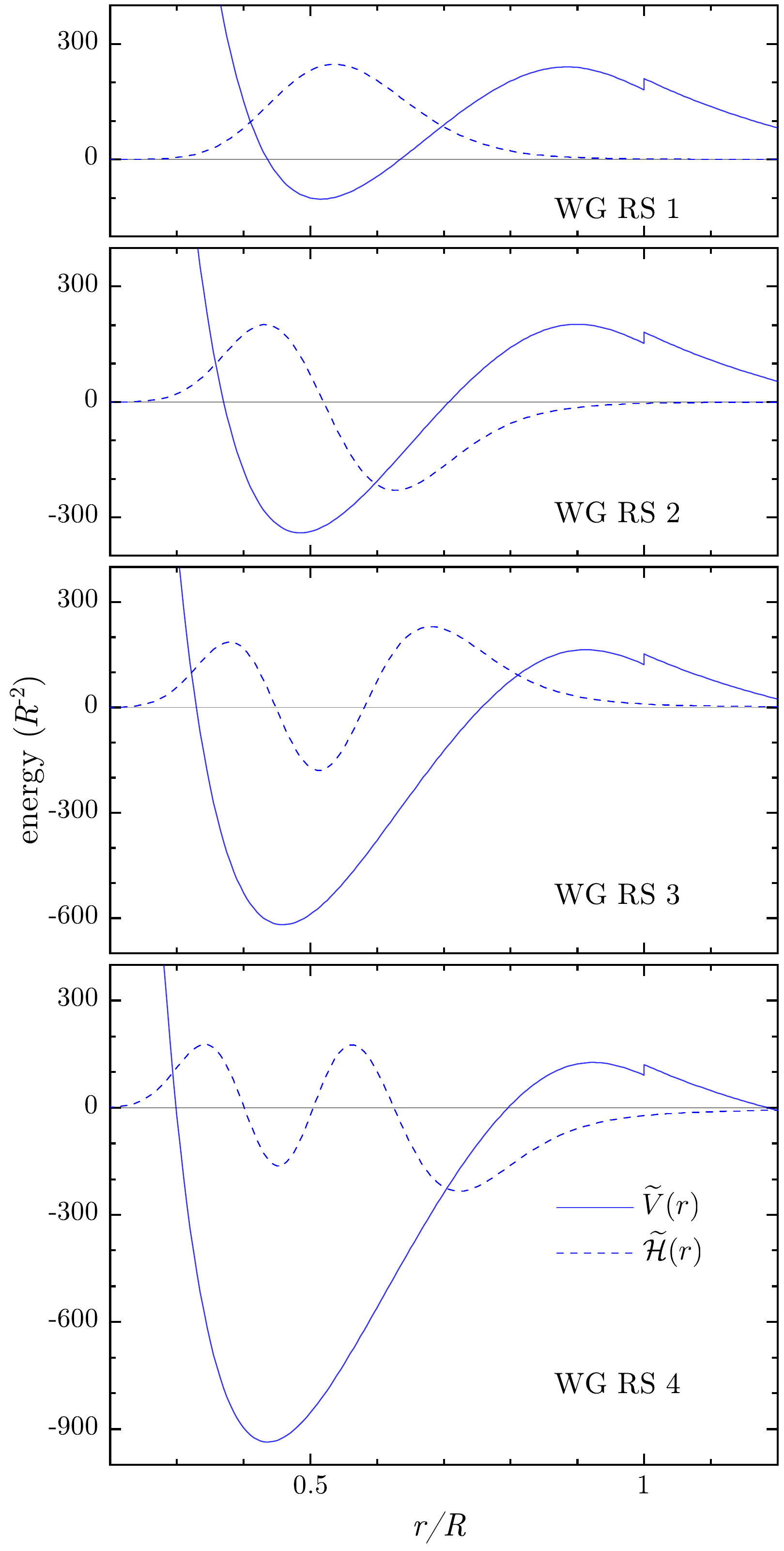}
	\caption{Real part of the effective potential $\widetilde  V^{\rm TM} $ and field  $\widetilde{\cal H} $ for the first four TM WG RSs, for the quadratic permittivity profile and $l=20$ as in \Fig{f:modes_and_potential}. WG RS 1 $k = 14.4 - 6.74\times10^{-9}i$, WG RS 2 $k = 15.4 - 3.51\times10^{-7}i$, WG RS 3 $k = 16.3 - 8.47\times10^{-6}i$, WG RS 4 $k = 17.2 - 1.22\times10^{-4}i$.}
\label{f:new_potential_and_excited_states}
\end{figure}

Following \cite{LockJQSRT17}, we introduce a substitution  $\mathcal{H}(r) = \sqrt{\varepsilon (r)} \widetilde{\mathcal H}(r)$, from which we find
\begin{align}
\dv[]{\mathcal{H}}{r} &= \frac{1}{2} \frac{\varepsilon'}{\sqrt \varepsilon} \widetilde{\mathcal H} +\sqrt \varepsilon \widetilde{\mathcal H}'\,, \\
\dv[2]{\mathcal{H}}{r} &= \left(-\frac{1}{4} \frac{ (\varepsilon')^2}{\varepsilon^\frac{3}{2}} + \frac{1}{2} \frac{ \varepsilon''}{\sqrt \varepsilon} \right)  \widetilde{\mathcal H} +  \frac{\varepsilon'}{\sqrt \varepsilon}  \widetilde{\mathcal H}' + \sqrt \varepsilon  \widetilde{\mathcal H}''\,,
\end{align}
where the prime indicates the derivative with respect to $r$ and where we omit the dependencies on $r$ for brevity. Using these expressions the wave equation takes the form

\begin{equation}
	\label{eq:radial_equation_TM_alt}
\left(\dv[2]{}{r} - \frac{\alpha^2}{r^2} + k^2 \varepsilon  - \frac{3}{4} \left(\frac{\varepsilon'}{\varepsilon}\right)^2 + \frac{1}{2} \frac{\varepsilon''}{\varepsilon} \right)  \widetilde{\mathcal H} = 0\,,
\end{equation}
in which the first derivative of the wave function present in \Eq{eq:radial_equation_TM} has cancelled out, so that the corresponding effective potential $\widetilde  V^{\rm TM}$ given by \Eq{eq:TM_potential_alt} is independent of the wave function $\widetilde{\mathcal H}$. This comes at the cost of adding a term containing the second derivative of the permittivity to $\widetilde  V^{\rm TM}$. We show $\widetilde  V^{\rm TM} $ in \Fig{f:new_potential_and_excited_states} for the lowest four WG modes, for the quadratic permittivity profile described in \Sec{ss:quadratic_permittivity}. Overall, the potential has a shape similar to $V^{\rm TE}$ shown in \Fig{f:modes_and_potential}(f), apart from the step at the sphere surface due to the contribution from the second derivative of the permittivity which has a discontinuity. As in the TE polarization, the potential is getting deeper with the mode number, and its minimum is slightly shifting towards the center, as it is clear from \Fig{f:new_potential_and_excited_states}.

\section{Morse potential}
\label{a:morse}

The Morse potential is a non-parabolic potential with known analytical solutions for energy levels and corresponding wave functions, often used to describe the binding of diatomic molecules \cite{DahlJCP88}. We take the Morse potential in the form
\begin{equation}
V_{\rm M}(r) = D_e \left\{1 - \exp[-a(r-r_e)]\right\}^2
\end{equation}
where $D_e$ is the dissociation energy, $r_e$ is the position of the potential minimum, and $a$ is an inverse well width. The potential is zero at $r=r_e$ and approaches $D_e$ asymptotically with increasing $r$. The bound energy levels of a quantum particle with a mass $M=\hbar^2 / 2$ in this potential are $E_n =  -a^2(\lambda - n - 1/2)^2 + D_e$, where $\lambda = \sqrt{D_e}/a$ and $n=0,1,...$ with $n<\lambda - 1/2$.

We apply this potential here to find approximate solutions for WG modes in the QMA, given by \Eq{eq:radial_equation_TE} for TE and \Eq{eq:radial_equation_TM_alt} for TM polarization. To do so, we match the coefficients of the Taylor expansion of the Morse potential $V_{\rm M}(r)$ and the corresponding QMA potential $V(r)$ at their minimum $r_e$ up to third order. Matching the value at the minimum is achieved by adding the value $V(r_e)$ to the Morse potential and its eigenenergies:
\begin{equation}
\label{a:eq:morse_energy_levels}
E_n = -a^2(\lambda - n - 1/2)^2 +D_e +V(r_e) \,.
\end{equation}
The first derivative of both potentials is zero at the minimum and is matched automatically by construction.
We then determine $D_e$ and $a$ by matching the second and third derivatives, yielding
\begin{equation}
\label{a:eq:morse_parameters}
V''(r_e)=2a^2 D_e  \quad \mbox{and} \quad V'''(r_e)=-6a^3 D_e\,,
\end{equation}
where the prime denotes the derivative with respect to $r$. As $V(r)$ depends on $k$, each WG mode has its own Morse potential parameters.

Now, since the solution corresponding to the WG mode has zero energy in the QMA, we can find an explicit equation determining the approximate value of the WG mode wavenumber $k_{\rm M}$. Eliminating $D_e$ and $a$ from  \Eqs{a:eq:morse_energy_levels}{a:eq:morse_parameters}, and requiring that $E_n=0$ yields
\begin{equation}
\label{a:eq:general_energy_levels}
\left[\frac{V'''}{3V''}\left(n+\frac{1}{2}\right)\right]^2 = V + \sqrt{2V''} \left(n+ \frac{1}{2}\right)\,,
\end{equation}
which is evaluated at $r=r_e$, where $r_e$ is determined by
\begin{equation}
\label{a:eq:minimum}
V'(r_e) = 0 \qquad \text{with} \qquad V''(r_e)>0\,,
\end{equation}
to select a minimum. Generally, \Eqs{a:eq:general_energy_levels}{a:eq:minimum} provide a nonlinear set of equations for $k_{\rm M}^2$, which can be solved numerically. Notably, for the case of a linear permittivity profile $\varepsilon(r)$ and TE polarization, the second and third derivatives of the potential are independent of $k$. They are given by $V''(r) = 6\alpha^2R^2/r^4$ and $V'''(r) = -24\alpha^2R^2/r^5$, so that the minimum position is determined by $r_e^3 = -2R\alpha^2 /(k^2\varepsilon')$. Inserting these into \Eq{a:eq:general_energy_levels} provides the explicit algebraic expression \Eq{eq:knTE_approximate} for the approximate wavenumbers of the WG modes.

\begin{table}[]
	\begin{tabular}{|c|c|c|c|c|c|c|}
		\hline
		$n$ & $k_{\rm RSE}R$ & $k_{\rm M}R$  & Relative error & $r_e/R$ & $1/aR$ & $D_e R^2$\\
		\hline
		0  & 54.11860 	& 54.12054 & 0.00004 	& 0.71708  & 0.53781 & 21266 \\
		1  & 55.26400 	& 55.27396 & 0.00018 	& 0.70707  & 0.53030 & 21872 \\
		2  & 56.40250 	& 56.42867 & 0.00046 	& 0.69739  & 0.52304 & 22484 \\
		3  & 57.53360 	& 57.58464 & 0.00089 	& 0.68802  & 0.51602 & 23100 \\
		4  & 58.65710 	& 58.74180 & 0.00144 	& 0.67896  & 0.50922 & 23721 \\
		5  & 59.77250 	& 59.90012 & 0.00214 	& 0.67018  & 0.50263 & 24347 \\
		6  & 60.87960 	& 61.05955 & 0.00296 	& 0.66167  & 0.49625 & 24977 \\
		7  & 61.97800 	& 62.22004 & 0.00391 	& 0.65341  & 0.49006 & 25612 \\
		8  & 63.06740 	& 63.38155 & 0.00498 	& 0.64541  & 0.48405 & 26251 \\
		9  & 64.14750 	& 64.54401 & 0.00618 	& 0.63763  & 0.47822 & 26895 \\
		10 & 65.21800 	& 65.70736 & 0.00750 	& 0.63008  & 0.47256 & 27544 \\
		11 & 66.27870 	& 66.87152 & 0.00894 	& 0.62275  & 0.46706 & 28196 \\
		\hline
	\end{tabular}
	\caption{Comparison of TE WG mode wavenumbers calculated by the RSE (real part) and the Morse approximation  \Eq{eq:knTE_approximate}, along with the Morse parameters for each fit. The relative error is calculated with respect to the RSE. Results are shown for the linear permittivity profile as in \Sec{ss:linear_permittivity} and $l=80$.}
	\label{t:k_morse}
\end{table}

\begin{table}
	\begin{tabular}{|c|c|c|c|c|c|}
		\hline
		& WG 1	& WG 2	& WG 3	& WG 4 & WG 5 \\
		\hline
		$ k_{\rm M}R$	& 54.12	& 55.26	& 56.40	& 57.53	& 58.66  \\
		\hline
		\hline
		$n$   			& \multicolumn{5}{c|}{$E_n(k_{\rm M})R^2$}  \\
		\hline
		0			&	0		&	-551	&	-1124	&	-1720	&	-2338	\\
		1			&	535	&	0		&	-558	&	-1139	&	-1742	\\
		2			&	1064	&	543	&	0		&	-566	&	-1154	\\
		3			&	1585	&	1080	&	551	&	0		&	-573	\\
		4			&	2100	&	1609	&	1095	&	558	&	0		\\
		5			&	2608	&	2131	&	1631	&	1109	&	565	\\
		\hline
	\end{tabular}
	\caption{Energy levels in the five different Morse potentials corresponding to the first five WG modes for TE polarization, $l=80$, and a linear permittivity profile as in \Sec{ss:linear_permittivity}.}
	\label{a:t:energy}
\end{table}
A fit of the effective potential $V(r)$ for the first WG mode ($n=0$) in TE polarization with a Morse potential $V_{\rm M}(r)$ is illustrated in \Fig{f:modes_and_potential}c, showing an excellent visual agreement between the two. Table~\ref{t:k_morse} shows a comparison of the WG mode wavenumbers calculated using the RSE with the approximate ones using the Morse potential, \Eq{eq:knTE_approximate}, revealing a high accuracy of the approximation with relative errors in the $10^{-3}-10^{-5}$ range.

Finally, \Tab{a:t:energy} shows the six lowest states in each of the Morse potentials corresponding to the first five WG modes in TE polarization. The state describing the WG mode has zero energy, and is changing from the first ($n=0$) to the fifth state ($n=4$) in the Morse potential. Importantly, the other states at non-zero energy are not describing WG modes, different from what could be implied by the QMA.

\section{Qualitative discussion of the TE-TM splitting and an example for $l=80$}
\label{a:tetm_degeneracy_l80}

\begin{figure}[!t]
	\includegraphics[width=1\linewidth]{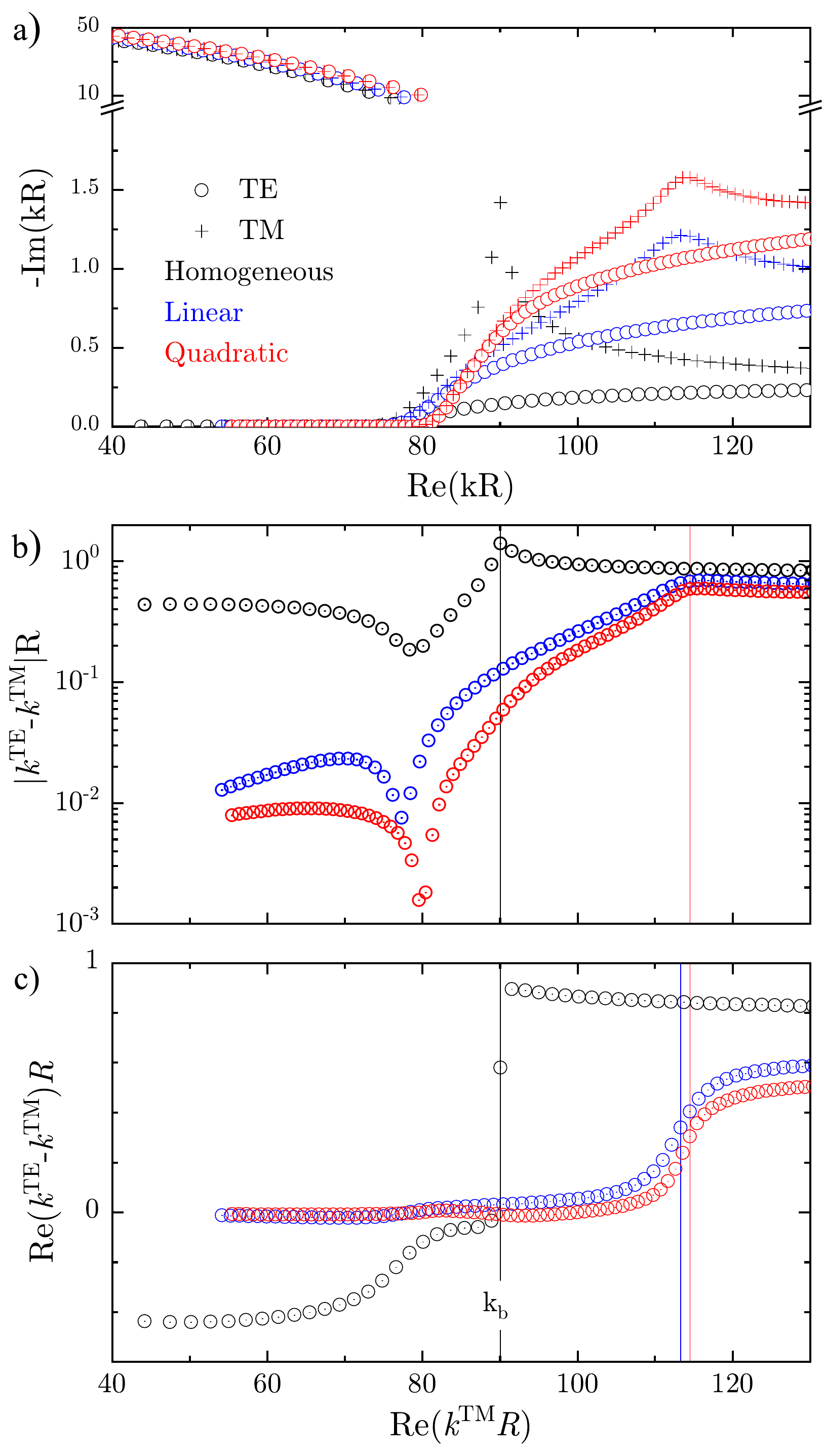}
	\caption{As \Fig{f:tetm_difference}, but for $l=80$.}
\label{f:tetm_difference_l80}
\end{figure}

One way to understand qualitatively the splitting of the fundamental WG (FWG) mode  for a given $l$ is to consider the RSs as light guided around the sphere as in a waveguide with an asymmetric cross-section in radial and polar direction. TE RSs have only the tangent direction of the electric field, while for TM RSs there is also a radial component. All modes have polar and radial confinement, both due to the spherical geometry. The radial confinement is determined by the effective potential as discussed in detail in \Sec{s:graded index}. As for the polar confinement, it is described e.g. for the FWG mode with $m=l$ by the analytic dependence of the field which is proportional to $\sin^l(\theta)$, where $\theta$ is the polar angle. We can find the angular width $\Theta$ of the polar confinement from the half maximum of the intensity  $\cos^{2l}(\Theta)=1/2$, after substituting  $\theta=\Theta+\pi/2$ into the above angular dependence of the field. This condition yields $\Theta=\pm\sqrt{\ln(2)/l}$ for $\l\gg1$. The full width at half maximum (FWHM) extension in polar direction \fwa\ is then approximately given by $\fwa=2\rp\sqrt{\ln 2 /l}$, with the peak radius $\rp$ of the RS, which for $l=20$ amounts to  about $0.37\,\rp$. For the constant permittivity (\Fig{f:modes_and_potential}b), we find $\rp\approx0.9\,R$, so that $\fwa\approx 0.33R$ and the FWHM in radial direction $\fwr\approx 0.15 R$. The RS asymmetry is thus about a factor of 2.2.
For the linear permittivity (\Fig{f:modes_and_potential}d), we find $\rp\approx0.71\,R$, so that $\fwa\approx 0.26R$ and the FWHM in radial direction $\fwr\approx 0.20 R$. The FWG mode asymmetry is thus about a factor of 1.3. For the quadratic permittivity (\Fig{f:modes_and_potential}f), we find $\rp\approx0.54\,R$, so that $\fwa\approx 0.20R$ and the FWHM in radial direction $\fwr\approx 0.17R$. The FWG mode asymmetry is thus about a factor of 1.2.  We see from these estimates that the RS asymmetry reduces when going from the constant to the linear and then further to the quadratic profile, and so does the TE-TM splitting.

To reduce the FWG mode asymmetry further, we have designed an index profile demonstrated and discussed in \Sec{ss:wide_well}. Looking at the FWG mode asymmetry in this case, we find $\rp\approx0.67\,R$, so that $\fwa\approx 0.25R$ and the FWHM in radial direction $\fwr\approx 0.33R$. The RS asymmetry is thus about a factor of 0.75, inverted compared to the other profiles. Still, the splitting has the same sign, showing that the FWG mode asymmetry is not a reliable predictor of the splitting. We note that as the field is extended in the radial direction the curvature of the sphere could be non-negligible, which is not taken into account in the asymmetry analysis.

In \Fig{f:tetm_difference_l80} the difference between the TE and nearest TM RSs for $l=80$ is shown, for the constant, linear, and quadratic permittivity profiles, using a basis size of $N=800$. The qualitative behaviour is similar to $l=20$ shown in \Fig{f:tetm_difference}, but the RSs are shifted to higher wavenumbers, and more WG RSs are present. The minimum splitting is reduced by approximately a factor of four, which is the increase factor of the tangent component of the wavenumber, $p=l/R$. This is due to the modes being more tightly packed, as can be seen from the approximate solution for linear profile based on the Morse approximation, \Eq{eq:knTE_approximate}, which contains a factor proportional to $ n/\alpha$, where $n$ is the mode number.

\section{RS separation}
\label{a:dispersion}

\begin{figure}[!t]
	\includegraphics[width=1\linewidth]{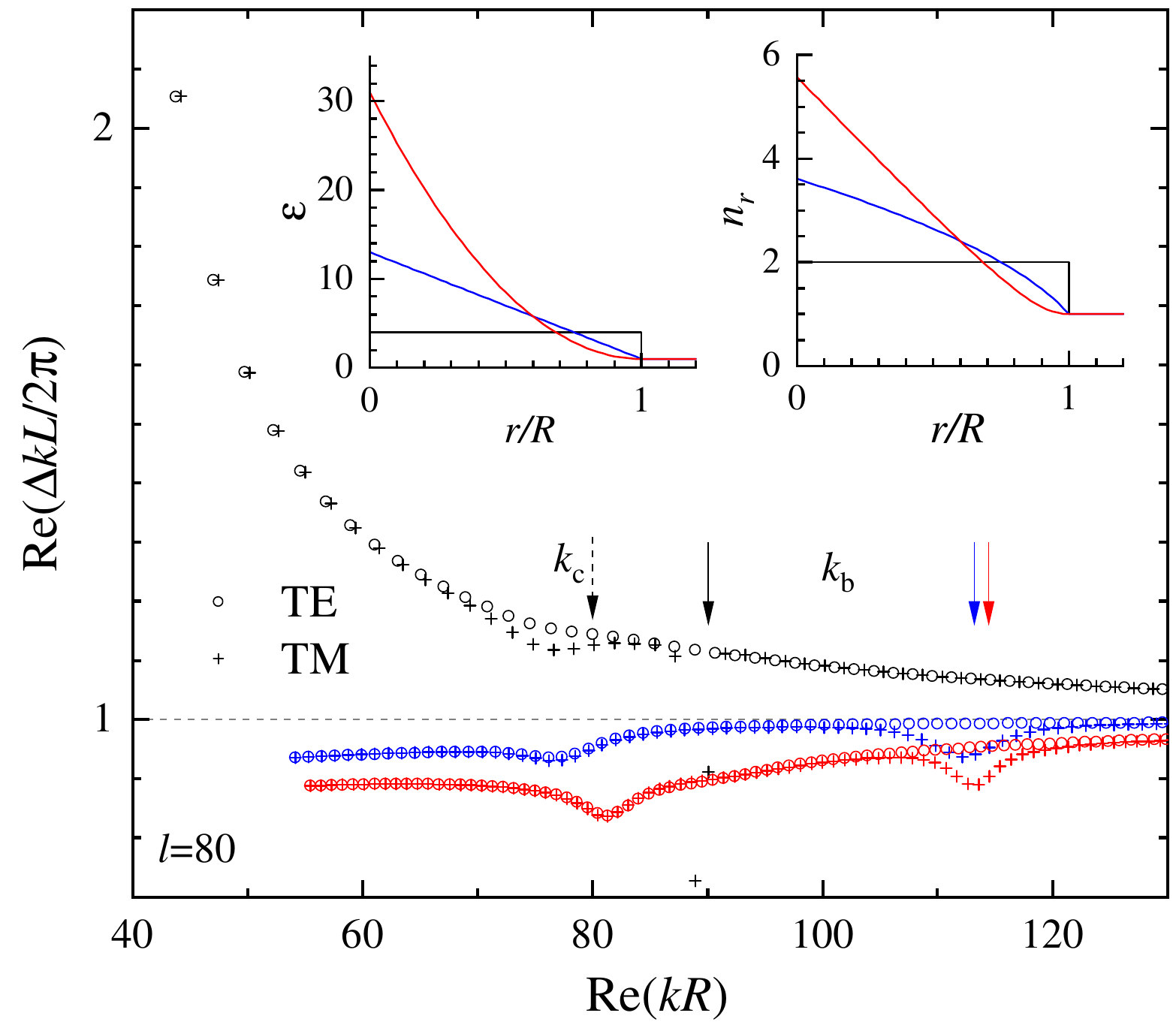}
	\caption{RS separation for $l=80$, for constant (black), linear (blue), and quadratic (red) permittivity profiles shown in the left inset. The right inset shows the corresponding refractive index profiles. The separation $\Delta k$ is taken between a RS at $k$ and the following RS of the same polarization. The vertical arrows indicate positions $k_b$ of the Brewster peak mode and $k_c$ of the critical angle of total internal reflection.}
	\label{f:dispersion}
\end{figure}

It is interesting to investigate the RS separation of each polarization for the different permittivity profiles, shown in \Fig{f:dispersion}. Let us consider in the ray picture a nearly normal incidence, corresponding to RS wavenumbers much larger than the critical wavenumber, $k_c=l/R$. In this case, the mode separation $\Delta k$ can be evaluated from the optical path length between successive reflections, $L = 2\int_0^R \sqrt{\varepsilon(r)} \dd r$.  $L/R$ takes the values of $4.0$,  $5.1$, and $6.0$, for the employed constant, linear, and quadratic permittivity profiles, respectively. Using the resonator condition of constructive interference of waves, $2L=n\lambda_n$, where $n$ is a natural number, and the missing even states discussed in \Sec{ss:phase_analysis}, we find $\Delta k=2\pi/\lambda_{n+1}-2\pi/\lambda_{n}=2\pi/L$. Therefore, $\Delta k$ in units of $2\pi/L$ tends towards unity for large $k$, which can be observed in \Fig{f:dispersion}.

Overall, for the constant profile, the spacing reduces with Re\,$k$, while for the linear and quadratic profiles, the spacing is nearly constant, increasing only slightly. There are two regions of deviation from the monotonous behaviour, indicated by vertical arrows in \Fig{f:dispersion}. Firstly, at the Brewster peak $k_b$, where the spacings of TM RSs, which otherwise are nearly identical to the TE RSs, are reduced in order to accommodate the additional Brewster RS, as discussed in \Sec{ss:brewster}. Secondly, at the critical wavenumber of total internal reflection at the surface, $k_c=l/R$, where both TE and TM RSs show a slightly reduced splitting, somewhat more pronounced for the TM RSs, specifically for the constant permittivity.

\newpage
\bibliography{references,langsrv}

\end{document}